\documentclass{aa}  

\usepackage{threeparttable}
\usepackage{graphicx}
\usepackage{natbib}
\bibpunct{(}{)}{;}{a}{}{,} 
\usepackage{float}
 \usepackage{xspace}
\usepackage{subfigure}
 \usepackage{url}
\usepackage{color}
\usepackage{xcolor}

\usepackage{txfonts}

\newcommand{\angstrom}{\textup{\AA}\xspace}
\def\kms{{\text{km\,s}$^{-1}$}\xspace}
\def\Teff{{\rm T$_{\rm{eff}}$}\xspace}
\def\Lsun{{\rm L$_{\odot}$}\xspace}
\def\Rsun{{\rm R$_{\odot}$}\xspace}
\def\Msun{{\rm M$_{\odot}$}\xspace}

\def\halpha{{\rm H$\alpha$}\xspace}

\newcommand{\io}[2]{#1\,{\textsc{#2}}}
\newcommand{\gaia}{\textit{Gaia}\xspace}
\def\ergs{{\rm\,erg\,s$^{-1}$}\xspace}
\def\he2{{He~{\small II}}\xspace}

\def\objname{AT\,2018bwo\xspace}

\def\HST{{\it HST}}

\begin{document}

\title{The luminous red nova AT\,2018bwo in NGC\,45 and its binary yellow supergiant progenitor \thanks{Table 1 is only available in electronic form
at the CDS via anonymous ftp to cdsarc.u-strasbg.fr (130.79.128.5)
or via http://cdsweb.u-strasbg.fr/cgi-bin/qcat?J/A+A/}}
\titlerunning{The luminous red nova AT\,2018bwo in NGC\,45}
   \author{Nadejda Blagorodnova \inst{1}\thanks{VENI fellow \email{n.blagorodnova@astro.ru.nl}}  
          \and
          Jakub Klencki \inst{1}
          \and
          Ond\v{r}ej Pejcha \inst{2}
          \and
          Paul M. Vreeswijk\inst{1}
          \and
          Howard E. Bond\inst{3,4}
          \and
          Kevin B. Burdge\inst{5}
          \and
          Kishalay De\inst{5}
          \and
          Christoffer Fremling\inst{5}
          \and
          Robert D. Gehrz\inst{6}
          \and
          Jacob E. Jencson\inst{7}
          \and
          Mansi M. Kasliwal\inst{6}
          \and
          Thomas Kupfer\inst{8}
          \and
          Ryan M. Lau\inst{9}
          \and
          Frank J. Masci\inst{10}
         \and
          R. Michael Rich\inst{11}
          \fnmsep
          }

   \institute{Department of Astrophysics/IMAPP, Radboud University, Nijmegen, The Netherlands
         \and
             Institute of Theoretical Physics, Faculty of Mathematics and Physics, Charles University, V Holešovičkách 2, 180 00, Praha 8, Czech Republic
        \and
            Department of Astronomy and Astrophysics, Pennsylvania State University, University Park, PA 16802, USA
        \and
            Space Telescope Science Institute, 3700 San Martin Drive, Baltimore, MD 21218, USA
        \and
            Cahill Center for Astrophysics, California Institute of Technology, 1200 E. California Blvd., Pasadena, CA 91125, USA
        \and
            Minnesota Institute for Astrophysics, School of Physics and Astronomy, University of Minnesota, 116 Church Street SE, Minneapolis, MN 55455, USA
        \and
            Steward Observatory, University of Arizona, 933 North Cherry Avenue, Tucson, AZ 85721-0065, USA
        \and
            Department of Physics and Astronomy, Texas Tech University, PO Box 41051, Lubbock, TX 79409, USA
        \and
            Institute of Space and Astronautical Science, Japan Aerospace Exploration Agency, 3-1-1 Yoshinodai, Chuo-ku, Sagamihara,
            Kanagawa 252-5210, Japan
        \and
            Caltech/IPAC, Mailcode 100-22, Pasadena, CA 91125, USA
        \and
            Department of Physics and Astronomy, The University of California, Los Angeles, CA 90095, USA
             }


   \date{Received 10 February 2021/ Accepted 10 June 2021}
   

 \abstract{Luminous red novae (LRNe) are astrophysical transients associated with the partial ejection of a binary system's common envelope (CE) shortly before its merger. Here we present the results of our photometric and spectroscopic follow-up campaign of \objname (DLT\,18x), a LRN discovered in NGC 45, and investigate its progenitor system using binary stellar-evolution models. The transient reached a peak magnitude of $M_r=-10.97\pm 0.11$ and maintained this brightness during its optical plateau of $t_p = 41 \pm 5$\,days. During this phase, it showed a rather stable photospheric temperature of $\sim$3300\,K and a luminosity of $\sim 10^{40}$\,\ergs. Although the luminosity and duration of \objname is comparable to the LRNe V838\,Mon and M31-2015LRN, its photosphere at early times appears larger and cooler, likely due to an extended mass-loss episode before the merger. Toward the end of the plateau, optical spectra showed a reddened continuum with strong molecular absorption bands. The IR spectrum at +103\,days after discovery was comparable to that of a M8.5\,II type star, analogous to an extended AGB star. The reprocessed emission by the cooling dust was also detected in the mid-infrared bands $\sim$1.5\,years after the outburst. Archival \textit{Spitzer\/} and {\it Hubble Space Telescope\/} data taken $10-14$\,years before the transient event suggest a progenitor star with $T_{\rm{prog}}\sim 6500$\,K, $R_{\rm{prog}}\sim 100$\,\Rsun, and $L_{\rm{prog}}=2 \times 10^4$\,\Lsun, and an upper limit for optically thin warm (1000\,K) dust mass of $M_d < 10^{-6}$\,\Msun. Using stellar binary-evolution models, we determined the properties of binary systems consistent with the progenitor parameter space. For \objname, we infer a primary mass of 12$-$16\,\Msun, which is 9$-$45\% larger than the $\sim$11\,\Msun obtained using single-star evolution models. The system, consistent with a yellow-supergiant primary, was likely in a stable mass-transfer regime with $-2.4 \leq \rm log (\dot{M}/M_\odot \,\rm{yr}^{-1}) \leq -1.2$ a decade before the main instability occurred. During the dynamical merger, the system would have ejected 0.15$-$0.5\,\Msun with a velocity of $\sim$500\,\kms. }

   \keywords{binaries: general -- novae, cataclysmic variables -- Stars: flare -- Stars: individual: AT\,2018bwo  -- Stars: winds, outflows  -- Stars: evolution            }

   \maketitle
\section{Introduction}

Over half of all stars in our Universe are born as binary or multiple systems, and the fraction increases above 70\% among the most massive stars \citep{Moe2017ApJS,Sana2012Sci}. Studies suggest that 30\% of them have already interacted with their companions via mass transfer and 10\% have merged \citep{deMink2014ApJ}. The merger process starts when one of the stars overfills its Roche lobe (Roche lobe overflow; RLOF), initiating an unstable mass transfer toward its companion. This process may culminate with both stars orbiting inside a shared non corotating gaseous layer, called the common envelope \citep{Paczynski1976,Ivanova2013Rev}. The less-massive component quickly spirals into the gravitational well of the primary star, transferring the angular momentum of the binary to the envelope. At the termination of this phase, part of the envelope is ejected, leaving either a more compact binary or a fully coalesced star. This phenomenon is key to our understanding of the following: stripped-envelope or Type~Ia supernova progenitors, cataclysmic variables, X-ray binaries, and gravitational-wave sources \citep{IbenLivio1993PASP,Nelemans2000,Belczynski2016Natur,VignaGomez2020PASA}.

Mergers are a likely explanation for the existence of blue and red stragglers \citep{McCrea1964MNRAS,HillsDay1976ApL,Schneider2015ApJ,Britavskiy2019AA} and rapidly rotating spotted FK\,Com stars \citep{BoppStencel1981ApJ}. High-mass merger products might include magnetic stars \citep{Ferrario2009MNRAS,Schneider2019Natur}, the progenitor of supernova SN\,1987A \citep{MorrisPodsiadlowski2007Sci}, and luminous blue variables such as $\eta$~Car \citep{Mauerhan2013MNRAS}. 

Observationally, the merger process is thought to be accompanied by a peculiar type of outbursts called luminous red novae (LRNe). These events display peak luminosities lying between those of novae and supernovae and temperatures from $2000 - 10,000$\,K during peak brightness. Unlike novae, the outburst temperature quickly cools down as the envelope expands with $\sim$100\,\kms velocity, shifting into near-infrared (NIR) and mid-infrared (MIR) wavelengths and triggering dust formation.

Similar to the cold shells of Mira long-period variables, the ejected envelopes of LRNe become sites of formation of water vapor and metal oxides such as titanium oxide (TiO) and vanadium oxide (VO). For example, NIR spectroscopy of the Galactic stellar merger V838\,Mon at early times showed strong absorption bands corresponding to water, AlO, and \io{CO} (see \citealt{Banerjee2005ApJ}). These same components were observed in emission in another merger, V4332\,Sgr, $\sim$10 years after its outburst \citep{Banerjee2015ApJ}, revealing the slow condensation  of highly  under-oxidized SiO, AlO, Fe, and Mg dust grains in the outflow.

New discoveries by ongoing all-sky transient surveys and archival searches for LRN-like events have effectively doubled the number of known systems in the last couple years \citep{Pastorello2019a,Pastorello2019b,Cai2019AA,Stritzinger2020AA,Pastorello2020_1,Pastorello2020_2}. However, limited by the lack of archival high-resolution data on their host galaxies, the progenitors of LRNe remain largely unexplored. To date, reliable multiband detections of the systems in quiescence have only been reported for the following five LRNe outside of the Milky Way: M31-LRN2015 \citep{Dong2015,Williams2015}, M101\,OT2015-1 \citep{Blagorodnova2017ApJ}, SNHunt\,248 \citep{Mauerhan2018MNRAS}, AT\,2019zhd \citep{Pastorello2020_1}, and AT\,2020hat \citep{Pastorello2020_2}.

In this work, we present results of our follow-up campaign in the optical and NIR of the LRN \objname, and we interpret its progenitor system using binary stellar-evolution models. This study helps us better understand the evolutionary stage of the binary moments before the dynamical spiral-in phase and relate it to the outburst properties and its late-time evolution. This is the first time that a LRN progenitor has been studied in agreement with its binary nature.

\section{Observations}

\subsection{Discovery, host galaxy, and reddening}

\objname (DLT\,18x) was discovered at RA.=00:14:01.720, DEC.= $-$23:11:35.84 (J2000) and reported to the transient name server (TNS\footnote{\url{https://wis-tns.weizmann.ac.il/}}) by the Distance Less Than 40 Mpc survey \citep[DLT40;][]{Tartaglia2018ApJ} on UT 2018-05-22 22:16:19 with an unfiltered \textit{Clear} magnitude of 16.44 \citep{ATel11665,2018ATel11666}. The last nondetection 6\,days earlier with a limiting magnitude of \textit{Clear}=19.55 shows that the source was detected soon after the outburst onset. Subsequently, the transient was also reported by two other surveys: ATLAS \citep[ATLAS\,18qgb;][]{Tonry2018PASP} and {\it Gaia\/} Science Alerts \citep[Gaia\,18blv;][]{Hodgkin2013RSPTA,Hodgkin2021arXiv}. Although the source was classified on TNS as an Intermediate Luminosity Red Transient (ILRT) on 2018-05-23 18:00:01 by the extended Public ESO Spectroscopic Survey for Transient Objects \citep[ePESSTO;][]{Smartt2015AA}, its plausible classification as a LRN was not ruled out \citep{2018ATel11669}.

The host galaxy \object{NGC 45} has a spectroscopic redshift of $z=0.00156$ \citep{Springob2005ApJS}. The distances to NGC 45 reported in literature span a wide range of values from 4.47\,Mpc \citep{Bottinelli1985AAS} to 13.90 \citep{Willick1997ApJS}. Given the uncertainty, we choose the method that provides the most accurate individual distances, based on the tip of the red giant branch (TRGB). In this work, we adopt one of the most recent measurements of $D_L=6.64\pm0.10$\,Mpc \citep{Tully2013AJ}, corresponding to a distance modulus of $\mu=29.11\pm0.10$.

The foreground Milky Way extinction toward NGC 45 corresponds to $A_V=0.058$ \citep{Schlafly2011}, which translates into a reddening of $E(B-V)=0.0187$ for a nominal extinction law. The average extinction within the host galaxy was derived by \citet{Mora2007AA} by modelling young stellar clusters at different metallicities. The best representation of cluster colors was reached for a host-galaxy metallicity between $Z=0.006$ and 0.008 and a mean reddening of $E(B-V)=0.04$. In our work we assume this value to be an upper limit and adopt an extinction value of $A_V^{\rm host}=0.03^{+0.09}_{-0.03}$, motivated by the modelling of the progenitor SED in Sect.~\ref{sec:progenitor_modelling}. Thus, the total fiducial reddening used through this work is $E(B-V)=0.029$ ($A_V=0.089$).

\subsection{Optical and IR photometry}

\begin{figure*}[h]
\hspace{-0.1cm}
\includegraphics[width=\textwidth]{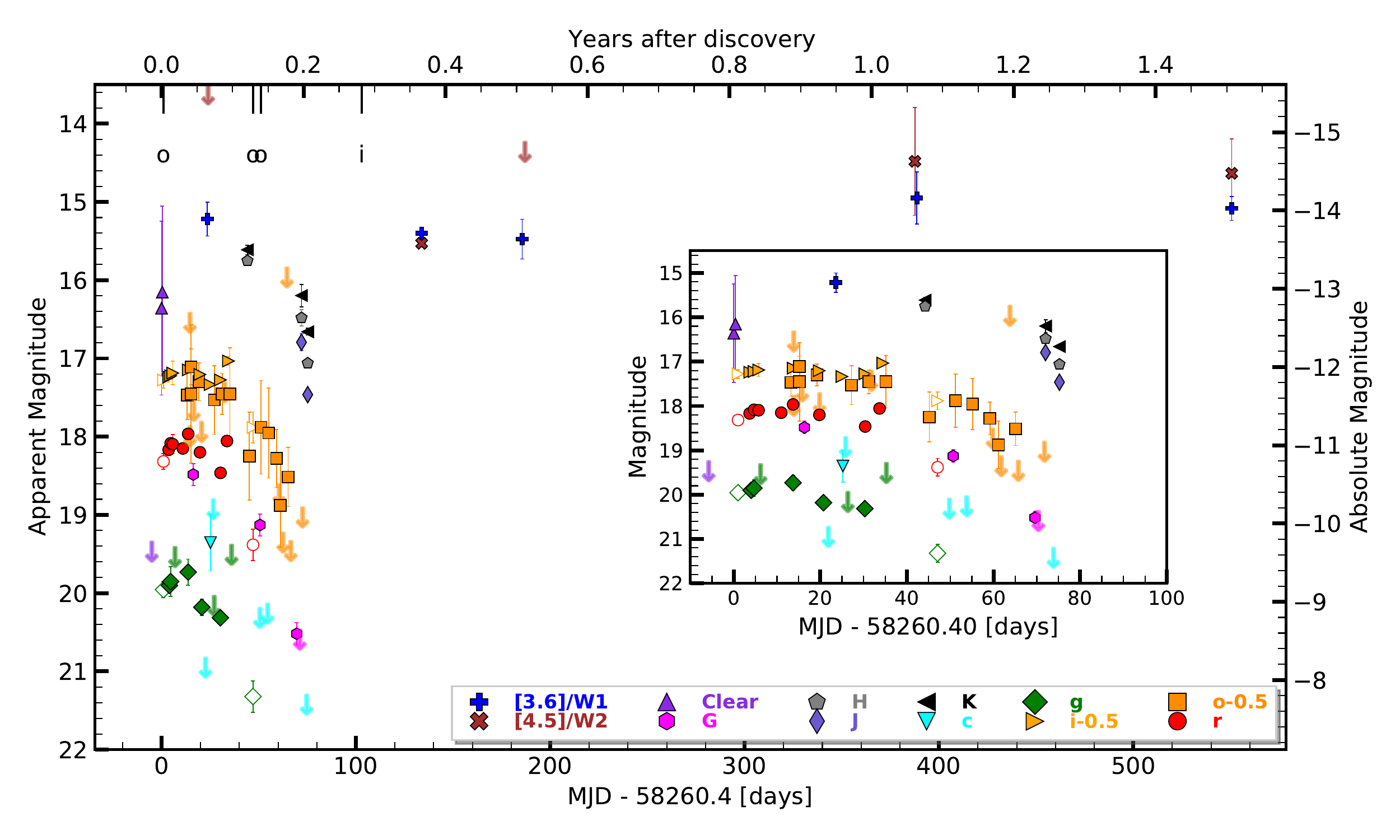}
\caption{Light curve of \objname. Optical magnitudes are in AB system and IR magnitudes are in Vega. The phase is shown as days from discovery. The open markers correspond to synthetic photometry from flux calibrated spectra. The ATLAS data are shown with a 1~day binning. The measurements are colour-coded and offset according to the legend. Arrows represent upper limits. The vertical lines at the top show the dates when optical (o) and infra-red (i) spectroscopy was obtained. The insert shows a zoom in of the light curve.}
\label{fig:lightcurves}
\end{figure*}

\begin{figure*}[h]
\centering
\includegraphics[width=\linewidth]{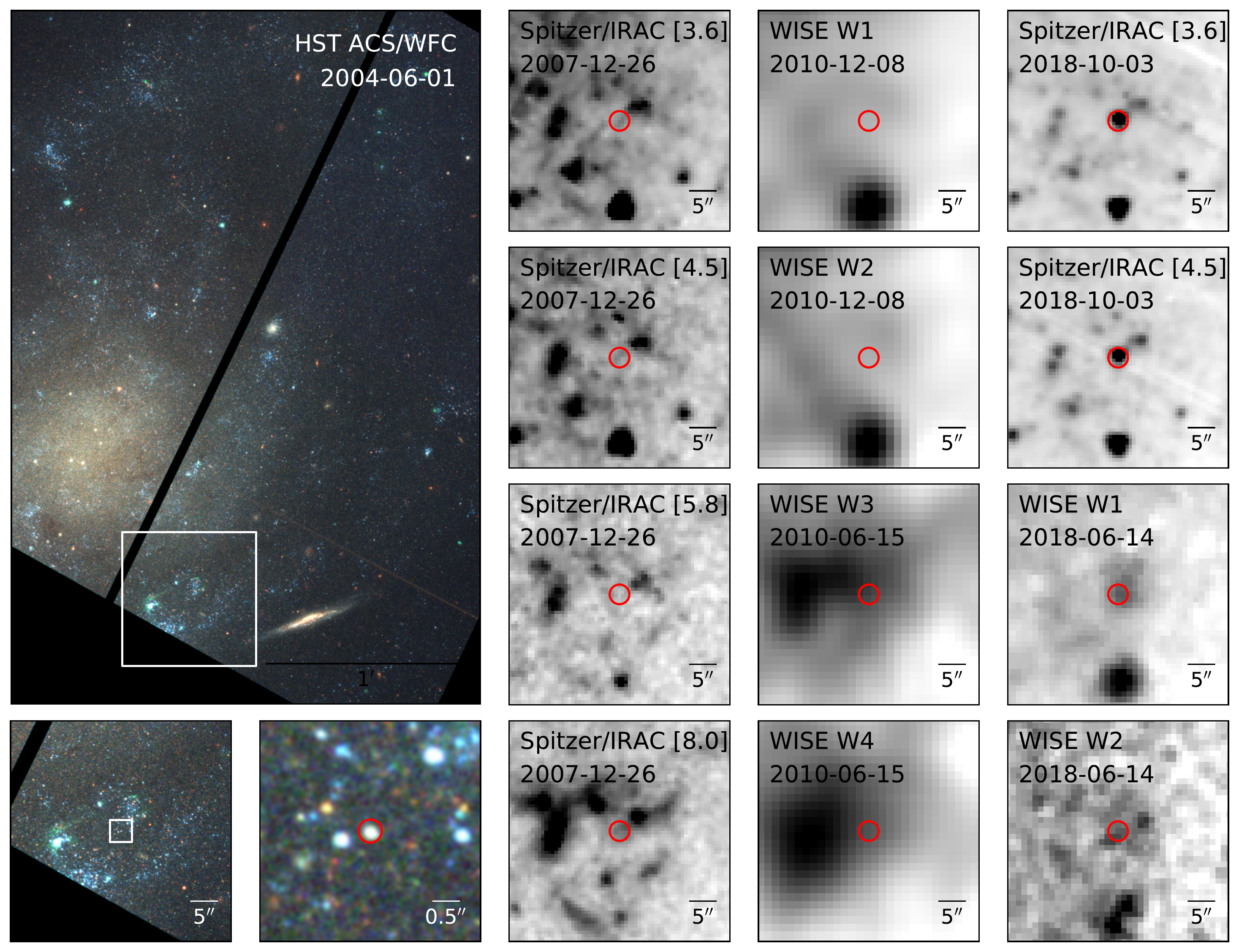} 
\caption{Left: Archival {\it HST\/} colour composite of the location of the host galaxy and the progenitor star. The zoom-in region is marked with a white square. The location of the progenitor is marked with a red circle. North is up and east is left. Middle: Archival \textit{Spitzer\/} and {\it WISE\/} imaging of the location of the progenitor. Right: Follow-up \textit{Spitzer\/} and {\it NEOWISE\/} imaging. }
\label{fig:progenitor}
\end{figure*}

Our follow-up campaign started soon after the transient was discovered and lasted up to 134\,days in the infrared (IR). Optical photometry for \objname was obtained using the Las Cumbres Observatory Global Telescope Network \citep[LCOGT;][]{Brown2013PASP} under the program NOAO2018B-008 (PI T. Kupfer). Once the transient had faded, at 250~days after discovery, we obtained the reference images of the field under the program NOAO2019A-011 (PI N. Blagorodnova). For each science image, we ran a custom adaptation of the image-subtraction algorithm \texttt{ZOGY} \citep{Zackay2016ApJ} and fit the field PSF to the residual image to obtain the photometry of the transient in the difference image.
In addition to our photometry, we included measurements from the time domain-surveys ATLAS, \gaia Science Alerts, and DLT40. The measurements from the ATLAS difference-image detection pipeline \citep{Smith2020PASP,Tonry2018ApJ}, reported in the orange $o$ and cyan $c$ filters, have been added with a binning of 1 day. The \gaia data are reported in the Vega system. The photometric data are provided in Table~\ref{tab:phot_ngc45} and Fig.~\ref{fig:lightcurves} shows the light curve of all combined photometric measurements. 

\renewcommand{\tabcolsep}{0.1cm}
\begin{table}
\begin{minipage}{1.\linewidth}
\begin{small}
\caption{Photometry of \objname. These magnitudes are not corrected for extinction. DLT40, LCO, and ATLAS magnitudes are in AB system. \textit{Gaia} and IR magnitudes are in Vega system. The full table is available as part of the online material. \label{tab:phot_ngc45}}
\begin{center}
\begin{tabular}{cccccccc}
\hline
MJD & Phase$^a$ & Telescope & Band  & Magnitude  & Comment   \\
    & (days) & &       & (mag) \\ \hline
58254.00 & -6.40 & Prompt5 & Clear &$>$19.55 &  [1] \\
58260.4 & 0.0 & Prompt5 & Clear & 16.44$\pm1.11$ & [1] \\
58260.8 & 0.4 & Prompt & Clear & 16.23$\pm1.10$ & [1] \\
58263.8 & 3.4 & LCO & r & 18.27$\pm0.04$ &  \\
58263.8 & 3.4 & LCO & i & 17.77$\pm0.05$ &  \\
58264.2 & 3.8 & LCO & i & 17.75$\pm0.03$ &  \\
58264.4 & 4.0 & LCO & r & 18.22$\pm0.04$ &  \\
58264.4 & 4.0 & LCO & i & 17.85$\pm0.03$ &  \\
58264.4 & 4.0 & LCO & g & 20.01$\pm0.08$ &  \\
58264.8 & 4.4 & LCO & g & 19.94$\pm0.09$ &  \\
... & ... & ... & ... & ... \\ \hline
\end{tabular}
\end{center}
$^a$The phase for \objname is relative to the discovery date MJD 58260.4.\\
Reference [1]: \citet{ATel11665}
\end{small}
\end{minipage}
\end{table}

Follow-up NIR photometry for \objname was obtained with the Wide-Field Infrared Camera on the Palomar 200-inch Hale telescope \citep[WIRC;][]{Wilson2003SPIE}, and the adaptive-optics instrument NIRC2 on Keck~2 \citep[LGS AO;][]{Wizinowich2006PASP}. The data were reduced using custom-developed pipelines, which first obtained individual frames using standard routines for dark subtraction and flat fielding. The individual dithered images were then combined for each filter. Due to the large field of view of the WIRC camera, the zeropoint for each observation night was estimated from field stars contained in the 2MASS \citep{2MASS_PSC_2006} catalog. For NIRC2, the zeropoints were estimated using two {\it Hubble Space Telescope\/} ({\it HST\/}) IR photometric standard stars, one of them observed at the beginning of the night and the other at the end.

MIR data were obtained with the Infrared Array Camera \citep[IRAC;][]{Fazio2004ApJS} on board the \textit{Spitzer} Space Telescope \citep[SST;][]{Werner2004ApJS,Gehrz2007RScI} as target of opportunity under the SPIRITS program \citep{Kasliwal2017ApJ}.  We used archival \textit{Spitzer} images taken on MJD 54460.228 ($\simeq$10.4 years before discovery) to run the SPIRITS difference-imaging pipeline on the photometry, taken 134\,days after discovery.

To complement our MIR photometry of \objname, we searched for archival data from the cryogenic Wide-Field Infrared Explorer mission \citep[\textit{WISE};][]{Wright2010AJ} and the Reactivation survey \citep[\textit{NEOWISE-R};][]{Mainzer2011ApJ,Mainzer2014ApJ}. Initially, the cryogenic \textit{WISE} mission covered the whole sky in the 3.4, 4.6, 12, and 22\,$\mu$m (W1$-$W4) from UTC January 2010 until UTC August 2010. Since its reactivation in December 2013, the \textit{NEOWISE-R} space mission has been scanning the sky in the W1 and W2 infrared bands with an approximate cadence of half a year. The latest \textit{NEOWISE} 2020 Data Release comprises observations between December 13, 2018 and December 13, 2019 UTC. 

The cadence of \textit{NEOWISE} provides $\sim$12 exposures taken within one or two consecutive days. To obtain higher signal-to-noise images, we initially coadded all the exposures obtained in the same ``observing block'' using the \texttt{ICORE} tool \citep{Masci2009ASPC}. A visual inspection at the location of the transient showed that no emission was present back in 2010, but a new source appeared in all the exposures obtained after the detection of the outburst in 2018 (see Fig.~\ref{fig:progenitor}). To compute the magnitude of this source, we performed aperture photometry following the instructions detailed in the \texttt{ICORE} manual. Unfortunately, the underlying region has non-negligible emission from the unresolved stellar population in the host galaxy, which can impact the accuracy of our measurements. Nevertheless, the agreement with our follow-up \textit{Spitzer} photometry, which was performed using a difference-imaging technique, shows that this impact is minor.

Our multiband light curve (see Fig.~\ref{fig:lightcurves}) shows that \objname quickly faded in optical magnitudes, while its IR counterpart remained bright for longer, similar to other LRNe. In the optical, the discovery epoch corresponded to a peak magnitude of $M_{\rm clear}=-12.9\pm1.1$\,mag, as reported by the DLT40 survey \citep{2018ATel11665}. However, there appears to be a systematic offset between DLT40 photometry and our difference imaging measurements (DLT40 team, private communication). Consequently, we decided to use our own data to derive the peak magnitudes, as we find that our photometry is consistent with the ATLAS survey and with the synthetic magnitudes obtained from the flux-calibrated ePESSTO classification spectrum taken at +1\,day (see open markers in Fig.~\ref{fig:lightcurves}). We compute our peak magnitudes from the photometric measurements obtained during the initial plateau phase, lasting 20 days after discovery. We compute average absolute magnitudes of $M_r =-10.97\pm 0.11$\,mag and $M_i=-11.40\pm 0.04$. Toward the end of the plateau, at day +34, we see a possible short secondary peak in the light curve in the $i$ and $r$ bands, although our cadence after the re-brightening is insufficient to characterize this feature. Our best approach for obtaining an additional photometric epoch consists of using the spectrum at +50.6\,days in order to derive the synthetic $g,r,i$-band photometry. In order to mitigate flux losses, we scale the spectrum to match the broadband \gaia $G$-band magnitude, obtained four days later.

\subsection{Spectroscopy}

Spectroscopic follow-up for \objname was done in both optical and NIR wavelengths. Although optical spectroscopy for LRNe has been covered in detail in the literature, this is the first study to present NIR data on an extragalactic LRN. The full log of the observations included in this work is provided in Table \ref{tab:speclog}, and the data are available through the online \texttt{WiseRep} repository \citep{YaronGal-Yam2012}. 

The optical \objname spectroscopic observations were conducted with the Low Resolution Imaging Spectrometer \citep[LRIS;][]{Oke1995PASP} on the Keck~I 10\,m telescope and with the Double Spectrograph \citep[DBSP;][]{OkeGunn1982PASP} on the Palomar 200-inch Hale telescope (P200). For LRIS we used a combination of grism 400/3400 and grating 400/8500, and a long 1$^{\prime\prime}$-wide slit, providing a resolution of $\sim$7\,\AA. For DBSP we used a combination of blue grating 600/4000 and red grating 316/7500 with a long 1.5$^{\prime\prime}$-wide slit, yielding a resolution of $\sim$10.5\,\AA. Finally, the PESSTO spectrum retrieved from TNS was observed with the EFOSC2 spectrograph mounted on the New Technology Telescope (NTT) in La Silla. Grisms \#11 and \#16 were used, with a resolution of 15.8 and 16\,\AA..

 The LRIS spectrum was reduced using the automated reduction pipeline in IDL \texttt{lpipe}\footnote{\url{http://www.astro.caltech.edu/~dperley/programs/lpipe.html}}, developed by D. Perley. The DBSP spectrum was reduced using the custom-developed DBSP pipeline \texttt{pyraf-dbsp} \citep{Bellm2016ascl.soft02002B}.

For NIR spectroscopy, we used the recently commissioned prism cross-dispersed Near-Infrared Echellette Spectrometer (NIRES) on Keck I\footnote{NIRES is a project developed at Caltech by PIs Keith Matthews and Tom Soifer.}. For this object, the data correspond to a single epoch taken at 4 months after discovery. The NIRES data was reduced using \texttt{NSX: NIRES Spectral eXtraction pipeline} \footnote{\url{http://www.astro.caltech.edu/~tb/nsx/}}. The data were corrected for telluric absorption using A0\,V-type stars at similar airmass and sky position. Two different calibration stars were observed, one at the beginning and one toward the end of the exposure. The flux was calibrated using the standard flux-calibration stars.

 \renewcommand{\tabcolsep}{0.1cm}
\begin{table*}
\begin{minipage}{0.9\linewidth}
\begin{small}
\caption{Log of spectroscopic observations of  \objname.}
\begin{center}
\begin{tabular}{rccccrc}
\hline
Phase$^a$ & MJD  & UTC & Telescope &     Slit & Exp  & Resolution \\ 
  (d) &  (d) &   &	+Instrument	& 	(arcsec) &	(s)	   & (km/s)  \\ \hline
+1.0 & 58261.4 & 2018-05-23 & NTT + EFOSC2  & 1.0 & 2000 & 730 \\
+47.1 & 58307.5 & 2018-07-08 & P200+DBSP & 1.5 & 900 & 480\\ 
+51.2 & 58311.6 & 2018-07-12 & Keck I+LRIS & 1.0   & 900 &  320\\
+103.1 & 58363.5 & 2018-09-02 & Keck I+NIRES &  0.55 & 3600 & 110\\ 
\hline
\end{tabular}
\end{center}
\label{tab:speclog}
\end{small}
\end{minipage}
\end{table*}

\section{Photometric and spectroscopic analysis}

\subsection{SED modelling}

\begin{figure}[h!]
\includegraphics[width=0.5\textwidth]{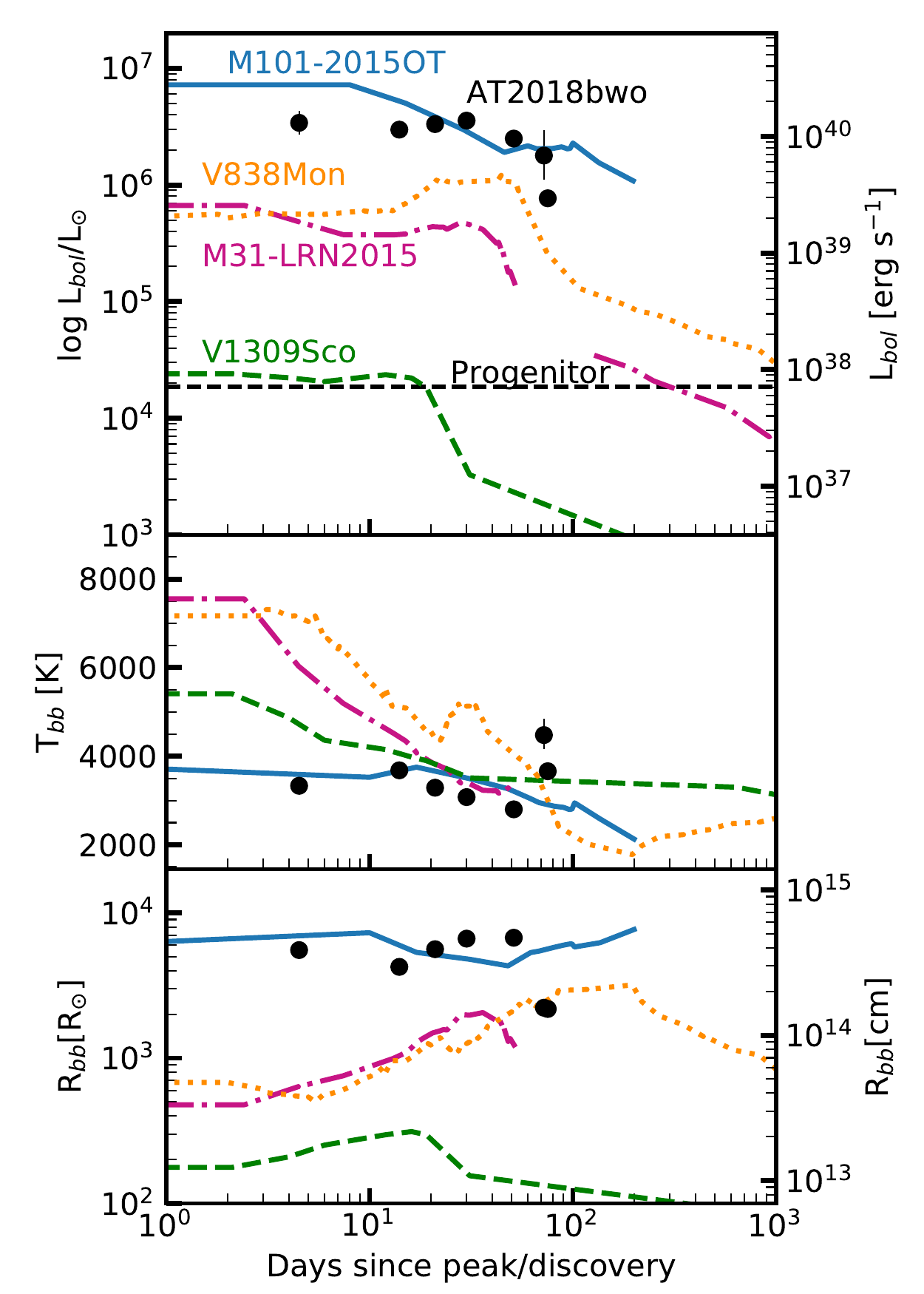}
\caption{Black body bolometric luminosity (top), temperature (middle) and radius (bottom) of \objname is shown with black circles. Coloured lines show the evolution of other LRNe V1309\,Sco \citep{Tylenda2011,Tylenda2016}, V838\,Mon \citep{Tylenda05a}, M31-LRN2015 \citep{MacLeod2017ApJ,Blagorodnova2020MNRAS} and M101\,OT2015-1 \citep{Blagorodnova2017ApJ}. The X-axis has been matched to the time of the first peak (second for M101-2015OT). The progenitor luminosity is shown by the thin dashed line.}
\label{fig:luminsity}
\end{figure}

The evolution of the outburst luminosity, temperature, and radius was estimated using black-body fits to the optical and infrared SEDs. The optical SEDs were obtained from LCO photometry for the epochs when all $g,r,i$ bands were available. The infrared SEDs were obtained when all $J,H,K$ or [3.6] and [4.5] bands were available. At +51\,days the best fit was performed on the flux calibrated Keck/LRIS spectrum. The fluxes were corrected for both Galactic extinction of $A_V=0.058$ and extinction in the host of $A_V=0.03$, in agreement with our best estimate for the progenitor (see Sect.~\ref{sec:progenitor_modelling}). When fitting the spectrum, we excluded the regions with stronger molecular absorptions. To estimate the fit uncertainties, the initial and final wavelength of the fitted region of the spectrum were shifted in steps of 100\,\angstrom within a 500\,\angstrom window. The results of our analysis are summarized in Table \ref{tab:luminosity}, and the LRN evolution is shown in Fig.~\ref{fig:luminsity}, along with other well-studied LRNe.

\begin{table}[]
    \centering
    \begin{tabular}{r c c c c}\hline
MJD & Phase &  T$_{\rm{eff}}$  & R$_{\rm{bb}}$ & L$_{\rm{bb}}$ \\
(day) & (day) &  (K) & (\Rsun) & (10$^6$\Lsun) \\ \hline
58264.9 & 4.5 & 3332$_{-71}^{+65}$ & 5542$_{-392}^{+439}$ & 3.41$_{-0.71}^{+0.88}$ \\ 
58274.4 & 14.0 & 3681$_{-51}^{+52}$ & 4245$_{-170}^{+181}$ & 2.98$_{-0.39}^{+0.45}$ \\ 
58281.4 & 21.0 & 3288$_{-25}^{+26}$ & 5619$_{-142}^{+147}$ & 3.33$_{-0.27}^{+0.29}$ \\ 
58290.4 & 30.0 & 3076$_{-21}^{+21}$ & 6644$_{-167}^{+171}$ & 3.56$_{-0.27}^{+0.29}$ \\ 
58311.6 & 51.2 & 2800$_{-12}^{+12}$ & 6740$_{-114}^{+114}$ & 2.50$_{-0.04}^{+0.04}$ \\ 
58332.4 & 72.0 & 4477$_{-312}^{+372}$ & 2222$_{-203}^{+208}$ & 1.79$_{-0.68}^{+1.16}$ \\ 
58335.6 & 75.2 & 3660$_{-77}^{+81}$ & 2179$_{-62}^{+63}$ & 0.77$_{-0.10}^{+0.12}$ \\ 
\hline
    \end{tabular}
    \caption{Best fit values and their 1$\sigma$ uncertainties for the SED of \objname at different epochs post discovery. In the fits, we fixed the amount of extinction to a total value of $A_V=$0.089.}
    \label{tab:luminosity}
\end{table}

The luminosity for \objname during the plateau is $\sim\!3\times 10^6$\,\Lsun ($\sim\!1.1 \times 10^{40}$ erg\,s$^{-1}$), which is a factor of 100 brighter than the progenitor. The total duration of the luminosity plateau is $\sim$70\,days, which is longer than observed in the optical wavelengths alone. Similar to other LRNe, the initial increase in luminosity during the plateau is followed by a sharp decline. 

At early times, the photospheric temperature of \objname  is among the lowest in the LRN sample, at only $3300\pm70$\,K. One other similar transient is M101\,OT2015-1, which showed a comparable range during its second peak. However, the temperature of its first peak was likely hotter, in agreement with the values inferred for other bright LRNe \citep{Pastorello2019a}. Around $\sim$70\,days after discovery, there is a fast increase in the temperature of the emission, along with the shrinking in the photospheric radius. The most likely explanation is that the most external layers become optically thin due to the expansion, and the photosphere moves inwards, where temperatures are higher.

\subsection{Spectroscopic characterization }

\subsubsection{Optical}

\begin{figure*}[h]
\includegraphics[width=0.95\textwidth]{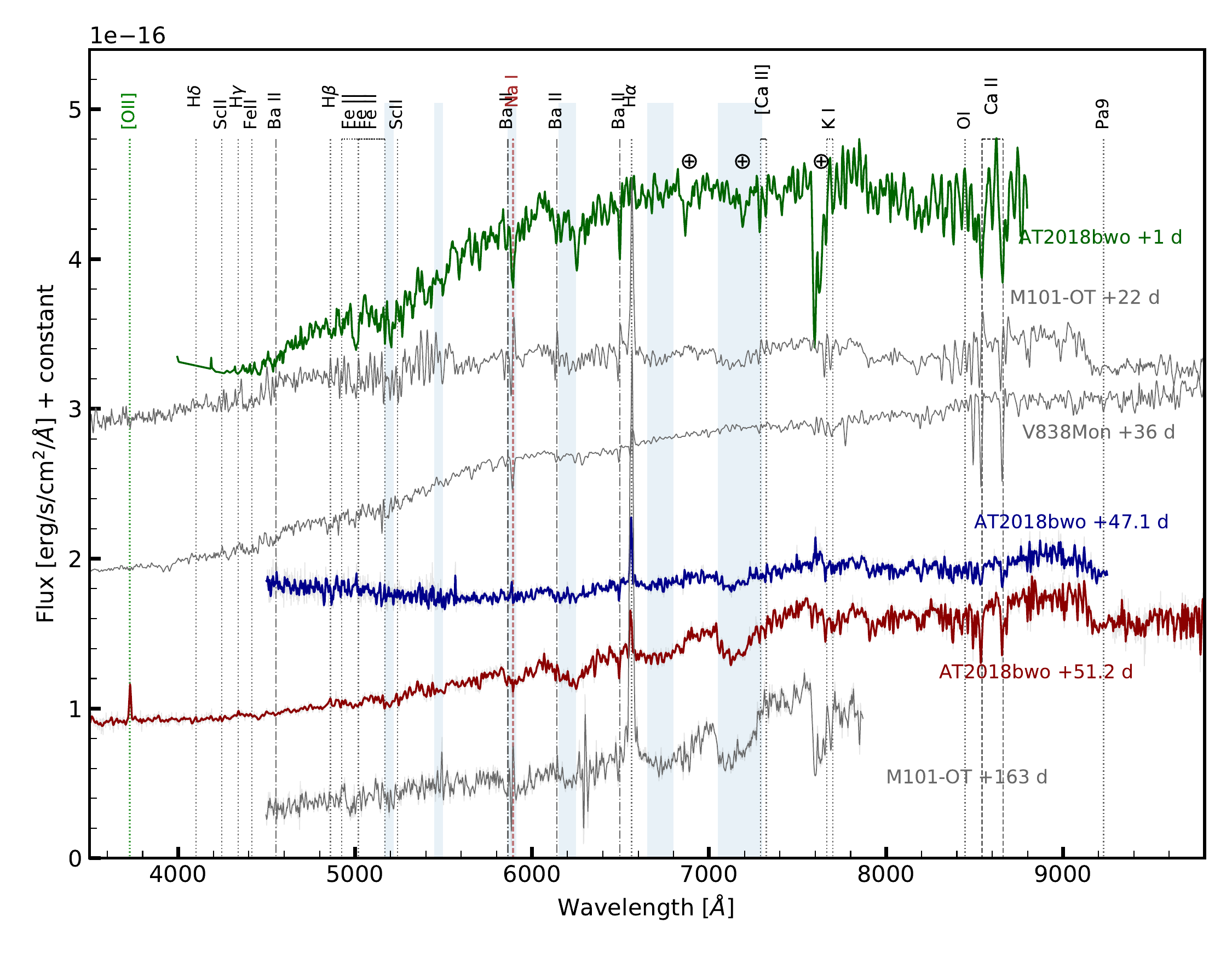} 
\caption{Spectral evolution of \objname (coloured lines) as compared to V838\.Mon \citep{Smith2016b} and M101\,OT2015-1 \citep{Blagorodnova2017ApJ} (gray lines). The main emission and absorption lines have been identified in the plot. The regions with strong molecular TiO absorption have been shown with light blue rectangles. Regions with strong telluric absorption are marked with the $\oplus$ sign. In these spectra, the [\io{Ca}{ii}] is detected in absorption, in contrast with the usual emission profile identified in other transients such as LBVs or ILRTs. }
\label{fig:spec_ngc45}
\end{figure*}

The optical spectral evolution of \objname is shown in Fig.~\ref{fig:spec_ngc45}. Once corrected for the redshift NGC\,45 (z=0.00156), in all three spectra we identify a blueshift of 270\,\kms for the major absorption lines, such as the \io{Ca}{ii} triplet, \io{K}{i} $\lambda\lambda$ 7665, 7699, \io{Ba}{ii}, and [\io{Ca}{ii}]. In addition, on top of the LRN spectrum, we also detect [\io{O}{ii}] $\lambda\lambda$3726, 3729 emission at the redshift of the host galaxy. Therefore, we attribute this line to the underlying \io{H}{ii} region where the star is located. The sky spectrum also shows an underlying \halpha emission at the rest wavelength of the host galaxy, which is offset from the peak of the \halpha line for the LRN. The presence of such background contamination may have an impact of some of our measurements (e.g., the P-Cygni profile identified at $-$800\,\kms may be due to oversubtraction).

The initial PESSTO classification spectrum, taken at +1\,day after discovery, reveals a cool photosphere, resembling a $\sim$3900\,K star. Analogous to the stellar spectra, the nova is dominated by absorption lines for low-ionization elements, such as \io{Ba}{ii}, [\io{Ca}{ii}], \io{Ca}{ii} IR triplet, and low-ionization states of \io{Fe}{ii} and \io{Sc}{ii}. This signature also resembles the early-time spectrum of V838\,Mon, taken at +36\,days after discovery (see comparison in Fig. \ref{fig:spec_ngc45}). Although we cannot identify any emission coming from Balmer transitions, the absorption for \halpha is not as strong as for an equivalent stellar type.

\begin{figure}[h]
\includegraphics[width=0.5\textwidth]{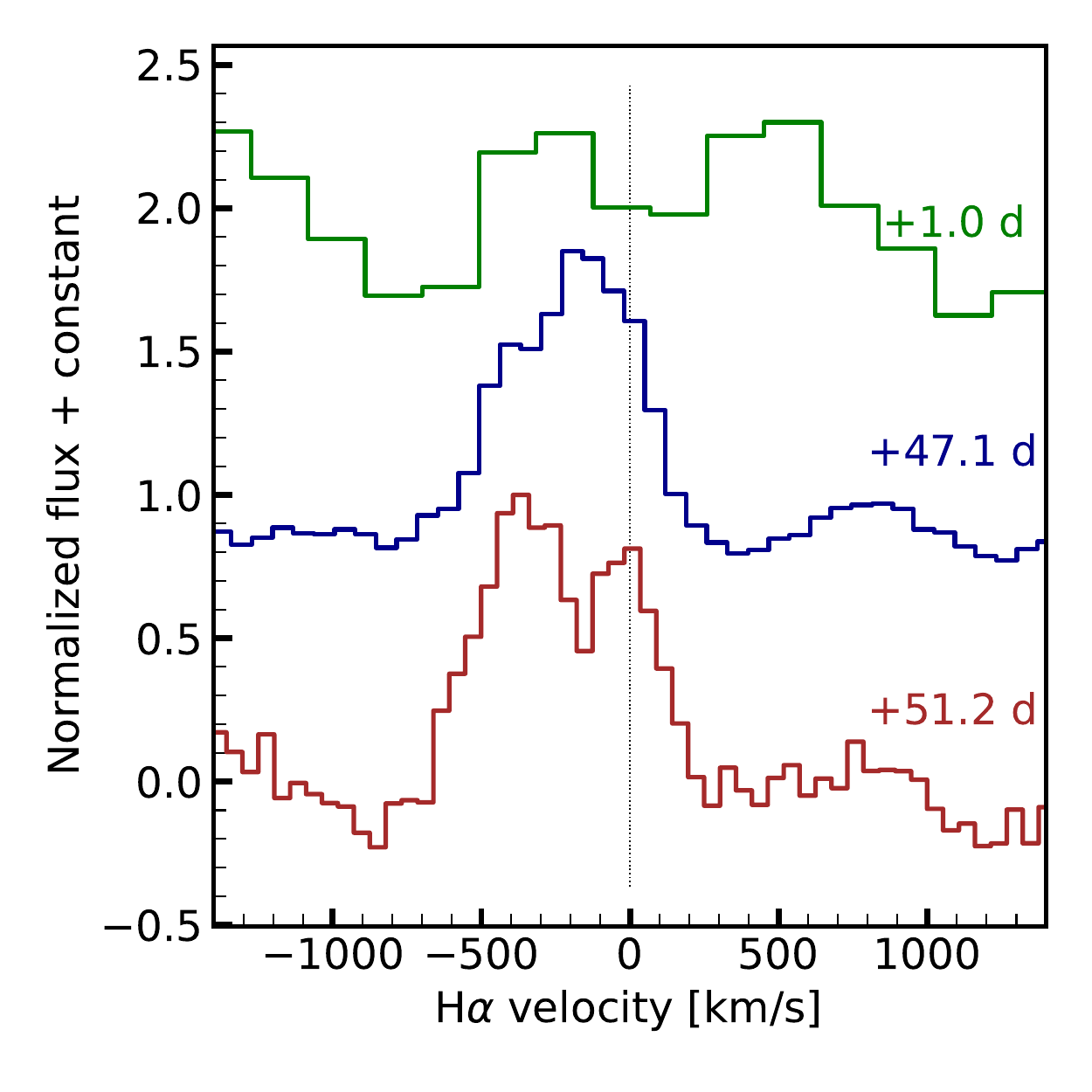} 
\caption{Normalized \halpha profile for the optical spectra. The line shows a likely P Cygni absorption at $\sim -800$\,\kms. Analogous to other LRNe, the emission line also presents a narrower absorption component at a lower velocity of $-150$\,\kms.}
\label{fig:halpha}
\end{figure}

The \halpha line shows signs of a possible P~Cygni profile at all three epochs, as displayed in Fig.~\ref{fig:halpha}. The expansion velocity derived from the minimum of the absorption for this line corresponds to $\sim$800\,\kms. However, we interpret this result with caution, as it could be an artifact from oversubtracting the galaxy background. The line also shows a bi-modal profile, likely associated with absorption in a colder expanding shell. At +47.1\,days, we derive an instrument-corrected FWHM of $450\pm18$\,\kms and $580\pm25$\,\kms for the later epoch at +51.2\,days. The approximate velocity of the absorption component for this last epoch is located at $-150$\,\kms with a FWHM of $\sim$150\,\kms. Similar profiles were observed for other Galactic LRNe, such as V4332\,Sgr \citep{Martini1999} and V838\,Mon \citep{Mason2010}, and the extragalactic outburst of M101\,OT2015-1 \citep{Blagorodnova2017ApJ}. 

Toward the end of the plateau, the last two spectra are already marked by the early appearance of molecular \io{TiO}{} and \io{VO}{} bands. The strength of the bands increases as the temperature of the spectrum drops to $\sim$2800\,K. As a comparison, molecular absorption was also detected in M101\,OT2015-1 as early as +22\,days after the secondary peak. For the brightest LRNe with longer plateaus, molecular absorption was detected at later stages, e.g., AT2018jfs \citep{Pastorello2019a} or NGC4490-2011OT1 \citep{Pastorello2019b}.

\subsubsection{Near-Infrared}

\begin{figure}[h]
\includegraphics[width=0.5\textwidth]{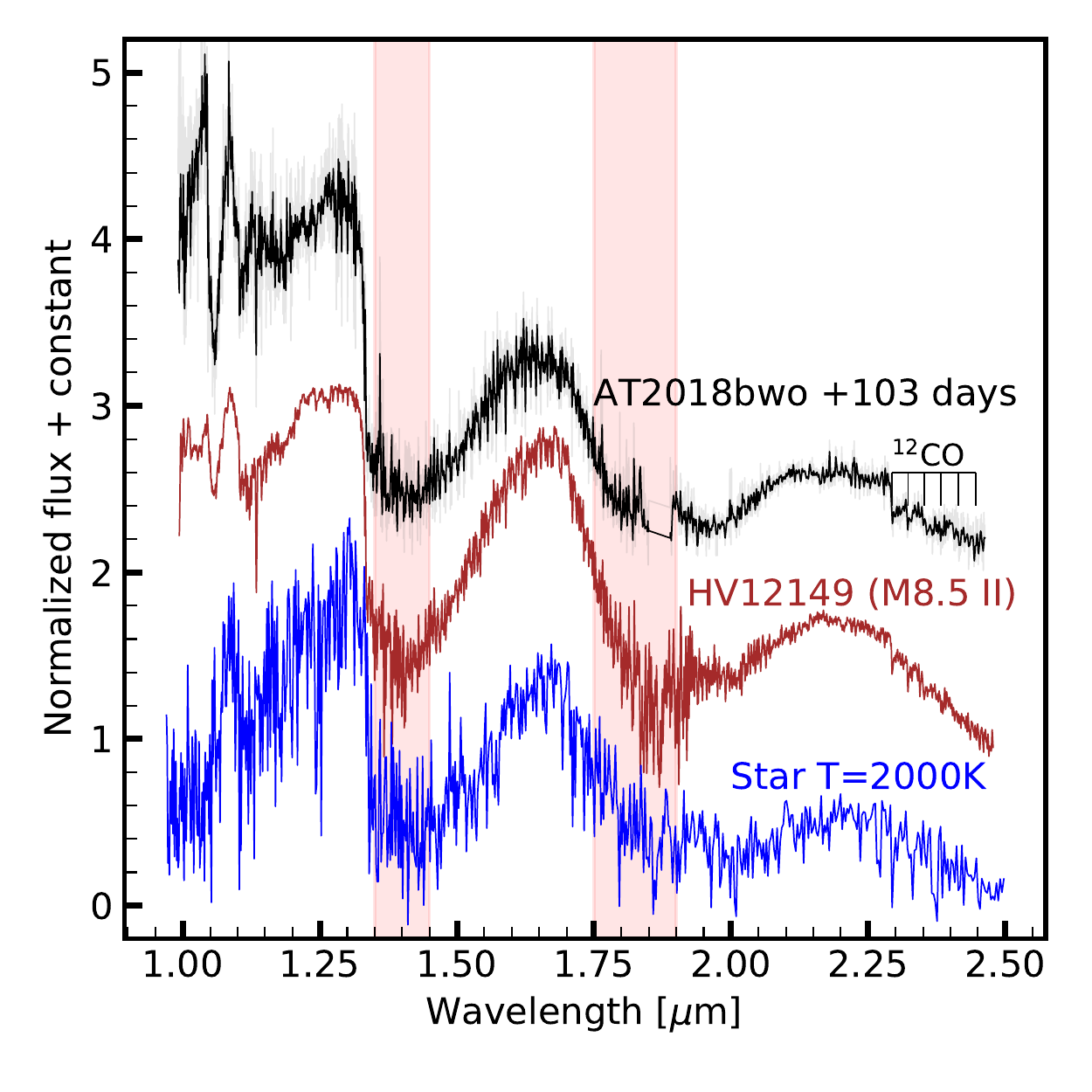} 
\caption{Spectrum of \objname at +103\,days after discovery (black line) as compared to a M8.5II type AGB star (brown) and a stellar model with \Teff=2000\,K (blue), solar metallicity and log(g)=0. The regions with strong water absorption are marked with shaded rectangles. For clarity, the spectra were smoothed with a Gaussian filter of 5\AA.  }
\label{fig:spec_ir_ngc45}
\end{figure}

\begin{figure*}[h]
\centering
\includegraphics[width=\textwidth]{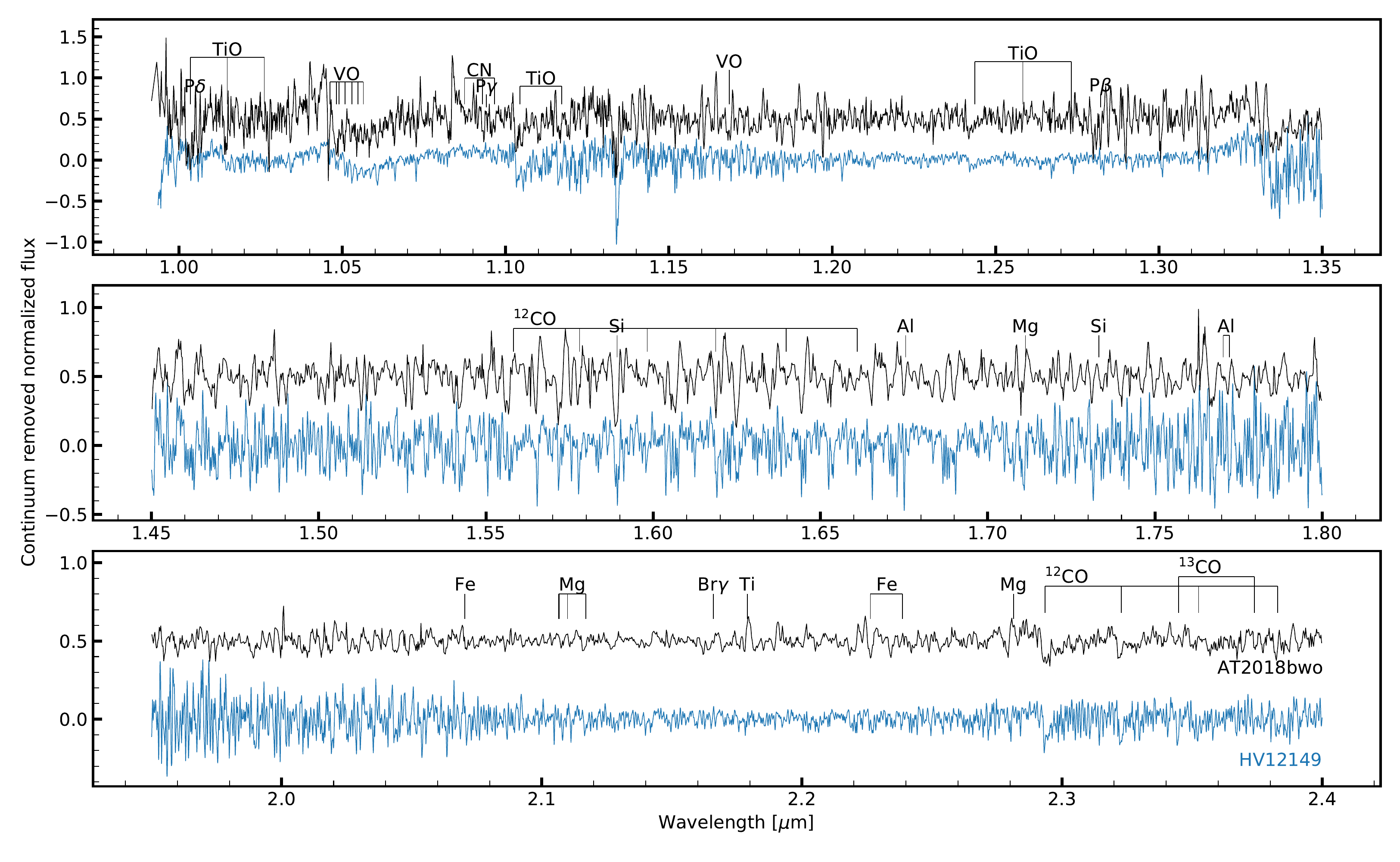} 
\caption{Continuum subtracted NIR spectra for \objname (black) and  a comparison star HV12149 (blue). Common IR molecular and atomic transitions are marked. The three panels correspond to $J$, $H$, $K$ spectra. Low signal-to-noise regions due to atmospheric extinction have been excluded. The coincidence of absorption features between the two spectra show that, albeit unidentified, most of the lines in the \objname are real. }
\label{fig:nir_nocontinuum}
\end{figure*} 

Fig.~\ref{fig:spec_ir_ngc45} shows the NIR spectrum of \objname at +103\,days, along with a stellar-atmosphere model and the spectrum of a cool giant star. The model was obtained from the Phoenix stellar spectral atlas \citep{Allard1995ApJ} and corresponds to a star with solar metallicity, $\log g=0$, and a temperature of 2000\,K, which is the lowest available temperature in their grid. The observed spectrum (HV 12149) was obtained from the DR2 of the X-shooter Spectral Library \citep[XSL;][]{Gonneau2020AA} and corresponds to a very luminous asymptotic-giant-branch (AGB) star, classified as
M8.5~II \citep{GonzalezFernandez2015AA}.

The classification of \objname at late times agrees with a late M-type giant star. The most prominent features are strong water vapor absorption bands around 1.1, 1.4, and 1.9\,$\mu$m, comparable to the ones detected in atmospheres of cool stellar types \citep{Lancon1992AAS}. These bands appear particularly noticeable in variable stars as compared to more static giants, as pulsations contribute to extending the stellar atmosphere \citep{LanconWood2000AAS}. Similar absorption troughs were also observed to appear at later times ($\sim$4 months after detection) in the spectrum of V838\,Mon \citep{Banerjee2002}, when the temperature of the emission dropped down to 2200$-$2400\,K. 

Fig.~\ref{fig:nir_nocontinuum} shows the continuum-subtracted spectrum of both the LRN and the cool AGB star. Molecular \io{TiO}{} and \io{VO}{} absorption features are detected around 1.05, 1.1,  and 1.25\,$\mu$m, in agreement with their identification in the optical spectrum. Due to the low gravity in the atmosphere of the LRN, the disappearance of molecular \io{TiO}{} and \io{VO}{} bands due to condensation will be pushed to extremely low temperatures, so that despite the temperature of the object being consistent with an L0-type star, we cannot classify the spectrum as such based on the strength of molecular absorption alone. However, early components of dust, such as \io{AlO}{} and \io{Al$_2$O$_3$}{} condense at higher temperatures than TiO. Therefore, their presence in the spectra of LRNe is not unexpected. In addition, we identified lines of CN in the $J$-band part of the spectrum. Similar to other cool giants, the $H$- and $K$-band spectra also show strong absorption from $^{12}$\io{CO}{} and $^{13}$\io{CO}{} bandheads\footnote{\url{https://www.gemini.edu/observing/resources/near-ir-resources/spectroscopy/co-lines-and-band-heads}}.

\section{Progenitor analysis} \label{sec:progenitor}

\subsection{Progenitor SED modelling} \label{sec:progenitor_modelling}

To measure the photometry of the progenitor of \objname, we searched the \HST\/ and {\it SST} archives. The progenitor location was imaged by the Advanced Camera for Surveys  Wide Field Channel (ACS/WFC) onboard {\it HST\/} on 2004 June~1, nearly 14 years before the detection of the merger outburst. The field was also imaged with {\it SST\/} on 2007 July 06 (PI Robert Kennicutt), about three years after the {\it HST\/} observations.

The precise location of the progenitor was identified by performing an astrometric registration of one of our follow-up NIRC2 adaptive-optics observations with the {\it HST}/ACS F814W frames. Our analysis shows an association within 1$\sigma$ between the location of the transient and a bright progenitor star in the {\it HST\/} images, shown in Fig.~\ref{fig:progenitor}. 

The photometry for the progenitor star was obtained from the analysis of NGC 45 stars presented in \citet{Mora2007AA}. The study performed aperture photometry with the \texttt{DAOPHOT} task in \texttt{IRAF}. Because of the greater flexibility of \texttt{DAOPHOT} to deal with variable background subtraction, we chose this approach over the magnitudes available in the Version 3 of the Hubble Source Catalog (HSCv3) \citep{HSC2016AJ}, which are based on aperture magnitudes from SExtractor V2.4.3 \citep{BertinArnouts1996}. The \texttt{DAOPHOT} magnitudes (reported in Vega system) are m$_{\rm{F435W}}$=23.878$\pm$0.038, m$_{\rm{F555W}}$=23.371$\pm$0.038, and m$_{\rm{F814W}}$=22.678$\pm$0.041, which are consistent with the \texttt{magaper2} values reported in HSCv3: m$_{\rm{F435W}}$=24.135, m$_{\rm{F555W}}$=23.608, and m$_{\rm{F814W}}$=23.346 (AB system), once corrected for aperture\footnote{\url{https://archive.stsci.edu/hst/hsc/help/HSC_faq/ci_ap_cor_table_2016.txt}}. The error bars for the HSCv3 magnitudes can be assumed as the average MAD in the archive for point sources of 23$-$24\,mag, with values between 0.04 and 0.05\,mag. 

We used the \texttt{pysynphot} \cite{pysynphot2013} software to convert the {\it HST\/} magnitudes into fluxes (corrected for extinction) and ran a custom Markov Chain Monte Carlo (MCMC) code \texttt{BBFit}\footnote{https://github.com/nblago/utils/blob/master/src/model/BBFit.py} based on \texttt{emcee} \citep{Foreman-Mackey2013PASP} to fit the black-body emission from the progenitor. Because the extinction in the host is unknown, we left it as a free parameter of the fit. In addition, we performed a second analysis using fixed values of $A_V$ for the host extinction. We applied corrective steps of 0.05\,mag to the $A_V$ parameter on top of the Milky Way value to retrieve the non-reddened fluxes. The maximum $A_V$ considered is 0.20\,mag, which is guided by the extinction values derived for young stellar clusters in NGC 45 \citep{Mora2007AA}. The best-fit results for the effective temperature, radius, and luminosity are reported in Table \ref{tab:progenitor}. 

In addition, we fit the progenitor fluxes using Kurucz model atmospheres \citep{Kurucz1993}. The models are available through the dusty radiative-transfer code \texttt{DUSTY} \citep{Ivezic97,Ivezic99,Elitzur01}. Following \citet{Adams15,Adams16,Adams17}, we used a MCMC wrapper around \texttt{DUSTY} to estimate the model parameters and their uncertainties. For the central source, we assumed a uniform prior on the temperature of $2500\leq T_* \leq 20000$\,K. The extinction along the line of sight was left as a free parameter. The results of this analysis, also reported in Table \ref{tab:progenitor}, agree with the fit using black-body emission.

\begin{figure}[hb] 
\includegraphics[width=0.5\textwidth]{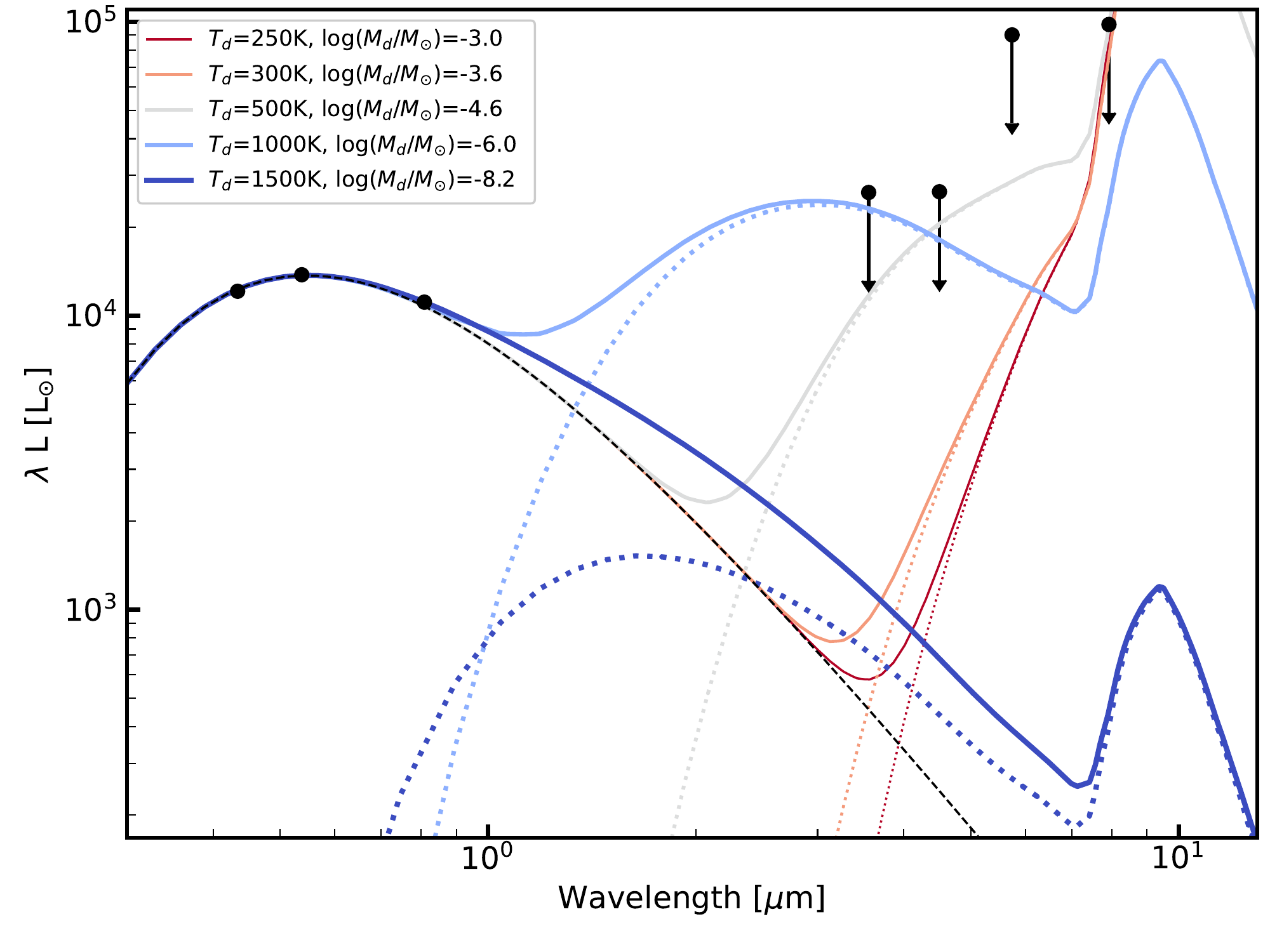} 
\caption{{\it HST\/} progenitor photometry and nondetection limits from \textit{Spitzer}. The dust is composed of $a=0.1\,\mu$m silicate grains radiating at a uniform temperature. The black line shows the contribution of the optical SED. The dotted lines show the contribution of the dust emission at different temperatures and the solid coloured lines represent the combined optical and infrared SED.}
\label{fig:dust_progenitor}
\end{figure}

The amount of preexisting warm dust in the progenitor system was estimated combining {\it HST\/} and \textit{SST\/} archival photometry. At 10.4\,years before the outburst, we inferred upper limits of $m_{[3.6]}>17.82$, $m_{[4.5]}>17.29$, $m_{[5.8]}>15.03$, and $m_{[8.0]}>13.88$\,mag from the IRAC mosaics. The MIPS 24$\mu$m data was not included in the analysis due to its poor resolution and contamination from nearby sources. Following the dust-modelling approach for the progenitor of M31-LRN2015 described in \citet{Blagorodnova2020MNRAS}---see their Sect.~3.3---we assumed that the dust is distributed in an optically thin shell around the system. The shell is formed by silicate or carbonaceous grains of $a=0.1\, \mu$m and radiates at a uniform black-body temperature. Fig.~\ref{fig:dust_progenitor} shows the maximum dust mass that would still be consistent with the nondetections, assuming silicate grains. Due to the wavelength range of \textit{SST} observations, the best constraints are obtained for warmer dust. For example, dust at $T_d=1500$\,K, if present, would need to have a mass of $M_d < 10^{-8.2}$\,\Msun. However, for colder dust at 1000\,K, 500\,K, and 250\,K, the upper limit increases to $M_d < 10^{-6.0}$, $M_d < 10^{-4.6}$, and $M_d<10^{-3.6}$\,\Msun, respectively. For graphite dust composition, the limits drop by almost one order of magnitude: $M_d < 10^{-9.4}$\,\Msun (1500\,K), $M_d < 10^{-6.8}$\,\Msun (1000\,K), and $M_d<10^{-5.2}$\,\Msun (500\,K).

In a similar analysis, we assumed a radial distribution of optically-thin dust corresponding to a stellar wind with $r^{-2}$ density profile and velocity of $50$\,km\,s$^{-1}$ \citep{kochanek11}. We find that the dust mass-loss rate is $\lesssim 10^{-7}\,M_\odot\,\text{yr}^{-1}$ to satisfy the \textit{Spitzer} limits. If we consider a relatively small extinction around the progenitor (as suggested by our optical SED modelling), a dusty wind along the line of sight is constrained to have dust mass-loss rate  $\lesssim 10^{-9}\,M_\odot\,\text{yr}^{-1}$ to remain optically thin. However, this constraint can be accommodated by confining the mass loss in the orbital plane of the binary and orienting the line of sight off the orbital plane.

 \renewcommand{\tabcolsep}{0.2cm}
\begin{table}[]
    \centering
\caption{Results of progenitor modeling using HST photometry obtained on MJD 53157.55.     \label{tab:progenitor}}
    \begin{tabular}{r r c c c}
Model & A$_V^{\rm{host}}$\,$^{\rm{a}}$ &  T$_{\rm{eff}}$  & R$_{\rm{bb}}$ & L$_{\rm{bb}}$ \\
& (mag) & (K) & (\Rsun) & (10$^4$\Lsun) \\ \hline

BB\,$^{\rm{b}}$ & 0.00 & 6441$_{-59}^{+62}$ & 108$_{-2}^{+2}$ & 1.81$_{-0.15}^{+0.15}$ \\
BB~~ & 0.05 & 6550$_{-62}^{+63}$ & 106$_{-2}^{+2}$ & 1.90$_{-0.16}^{+0.16}$ \\
BB~~ & 0.10 & 6650$_{-63}^{+66}$ & 105$_{-2}^{+2}$ & 1.98$_{-0.16}^{+0.16}$ \\
BB~~ & 0.15 & 6771$_{-68}^{+70}$ & 104$_{-2}^{+2}$ & 2.07$_{-0.18}^{+0.18}$ \\
BB~~ & 0.20 & 6894$_{-70}^{+72}$ & 103$_{-2}^{+2}$ & 2.17$_{-0.19}^{+0.19}$ \\ 
BB+$A_V$\,$^{\rm{c}}$ & 0.03$^{+0.09}_{-0.03}$ & 6708$_{-466}^{+545}$ & 101$_{-11}^{+14}$ & 1.86$_{-0.04}^{+0.13}$ \\
SSM $^{\rm{d}}$ & 0.03$^{+0.09}_{-0.03}$ & 6280$_{-300}^{+343}$ & 114$_{-9}^{+6}$ & 1.82$_{-0.08}^{+0.13}$ \\ \hline
    \end{tabular}
 \footnotesize{    \raggedleft{ $^{\rm{a}}$Amount of extinction in addition to the Milky Way. $^{\rm{b}}$Single black body model with a fixed value of extinction. $^{\rm{c}}$ Black body model with extinction as a free parameter. $^{\rm{d}}$ Fit using stellar spectral models. }}
\end{table}

\begin{figure}[hb]
\includegraphics[width=0.5\textwidth]{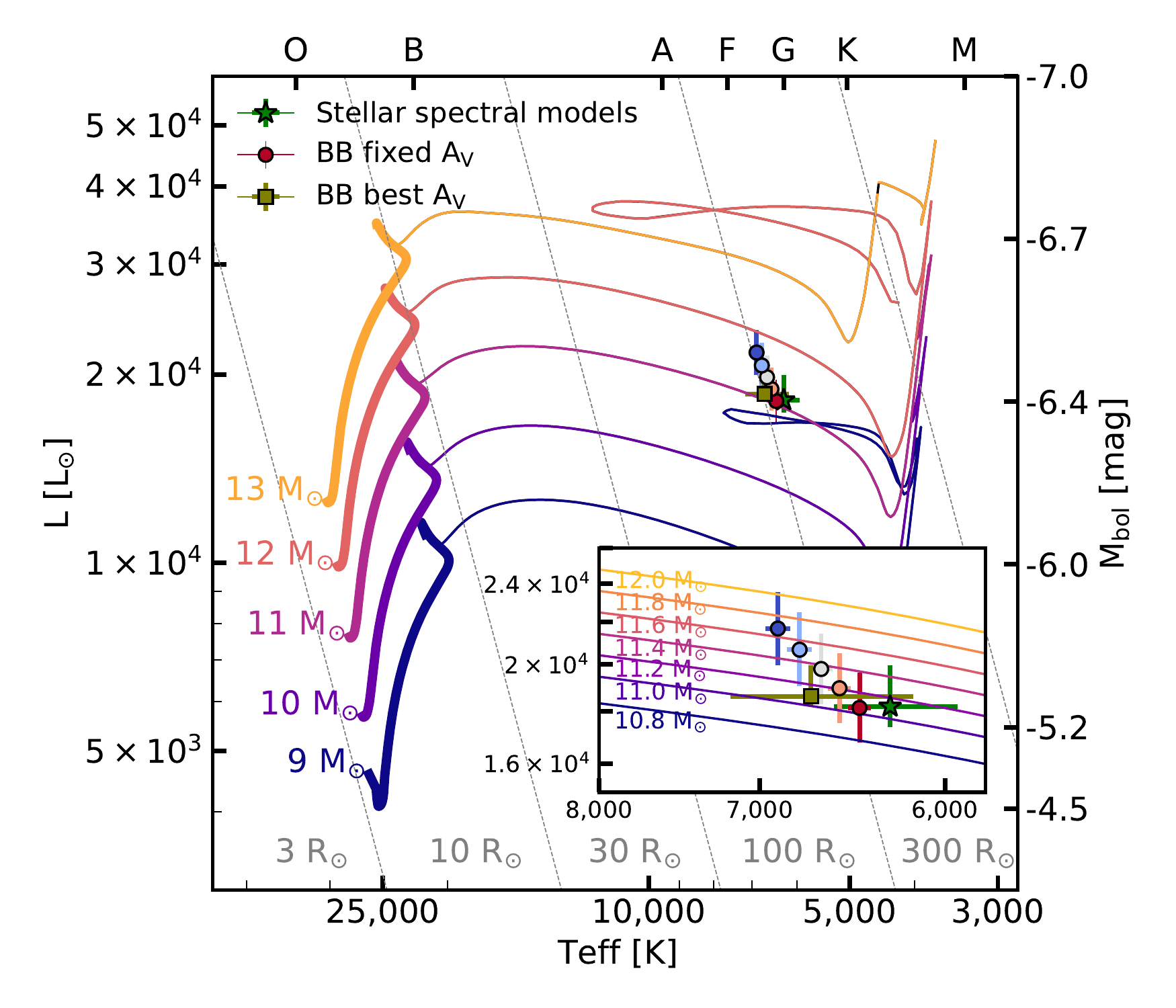}
\caption{Location of the \objname progenitor in the HR diagram. MESA single stellar evolution tracks are shown for stars between 9 and 13\Msun. The main sequence is marked with a thicker line. Different markers represent progenitor parameters using different models (see Table \ref{tab:progenitor}). The insert shows a zoomed in region around the progenitor location, along with a finer grid of models.}
\label{fig:mesa_tracks_single}
\end{figure}

\subsection{Single stellar evolution models}
\label{sec:single_stellar_models}

The progenitor was initially studied using single-star stellar evolution tracks from the code Modules for Experiments in Stellar Astrophysics \cite[MESA;][]{MESA2018ApJS}, available through the MIST \citep{MIST2016ApJ} web interface\footnote{http://waps.cfa.harvard.edu/MIST/index.html}. The retrieved stellar evolutionary tracks corresponded to a metallicity of $Z=0.0068$ ($\rm[Fe/H]=-0.32$), initial $v/v_{\rm crit}=0.4$. Our choice of metallicity was guided by the results of \citet{Mora2007AA}, who found that values between $Z = 0.006$ and 0.008  better represented the colors of young stellar clusters in NGC 45. Such a metallicity is between the values for the Small and Large Magellanic Clouds. 

The location of the progenitor, depicted in Fig.~\ref{fig:mesa_tracks_single}, can be associated with massive stars with initial masses between 11.0$-$12.0\,\Msun. The parameter space of the progenitor is consistent with the Hertzsprung gap: a fast transitioning phase between the main sequence and the red-giant branch. As shown in the bottom half of the figure, after leaving the main sequence, the radius of these stars expands considerably with age. For a lower-mass star around $\sim$9\,\Msun, the progenitor could also be consistent with being in a ``blue-loop'', that is, a star that had previously expanded to the red-giant branch and then shrunk again after the onset of core helium burning. However, provided that lower-mass stars would have expanded beyond the current radius at an earlier evolutionary stage, any interaction with a close companion would have occurred by that time. Therefore, we assume that the star is most likely in its first expansion after finishing core H burning, which theoretically is the most likely scenario for a stellar merger. In this case, assuming our fiducial extinction value of $A_V=0.089$, the most likely primary-star progenitor has a zero-age main-sequence mass of $M=11.0_{-0.2}^{+0.6}$\,\Msun. For this progenitor mass, the star would be $\sim$20.8\,Myr old and would have developed a He core with a mass of $M_{c}\simeq 2$\,\Msun, with an envelope mass of $M_{\rm env}\simeq8.80$\,\Msun.

\subsection{Binary stellar-evolution models }
\label{sec:binary_mesa_models}

The use of single stellar tracks to infer physical properties of the progenitor (Sect.~\ref{sec:single_stellar_models}) is only justified if at the time of observation 
the mass transfer was yet to commence. This requires that the timespan from the onset of RLOF to the onset of the CE phase is very short 
($<$10\,yr in the case of \objname). In the case of yellow-supergiant donors such as the \objname progenitor, such a scenario is only possible in rare cases of an extreme mass ratio ($q \gtrsim 20$, see discussion in Sect.~\ref{sec:disc_binary}). More likely, the mass transfer is driven by thermal-timescale expansion of the radiative envelope. As a result, it takes several hundreds of years 
before the mass-transfer rate rises to high values $\gtrsim 10^{-3}-10^{-2} \rm M_{\rm \odot} \, yr^{-1}$ \citep[e.g.,][]{Wellstein2001}
and, eventually, culminates in a runaway coalescence of the binary \citep{MacLeod2020ApJ_runaway}.
Therefore, the progenitor observed 10$-$13 yr before the LRN was most likely already transferring mass at a significant rate $\dot{M}$ onto 
its unseen companion.

\begin{figure}
{\includegraphics[width=0.5 \textwidth]{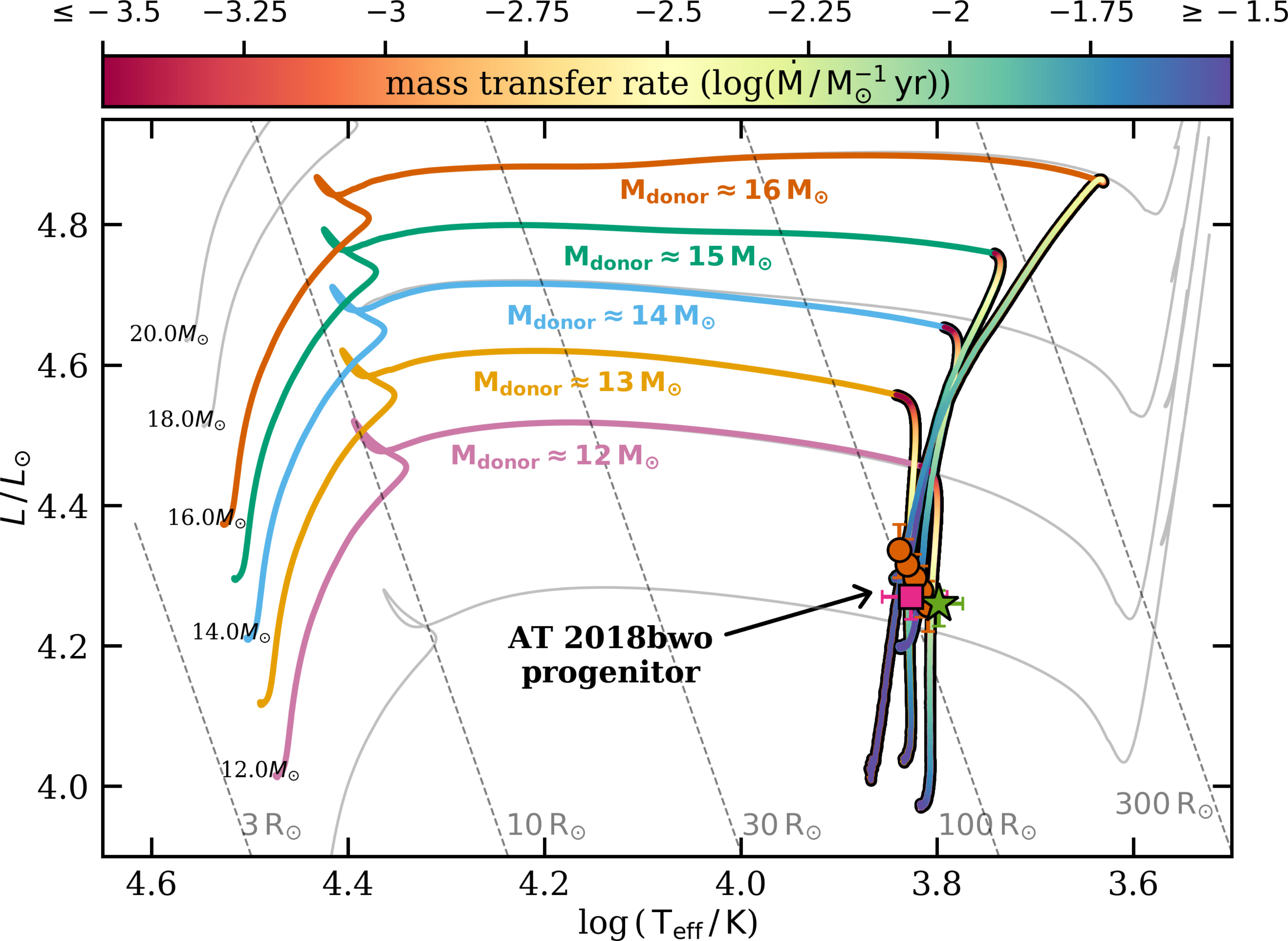} }
\caption{Evolution of donor stars of different masses in the HR diagram, 
derived from binary stellar evolution tracks computed with MESA. 
Nonrotating single stellar tracks from \citet{Klencki2020a} are plotted 
with thin gray lines. 
The binary models assume a mass ratio of $q = M_{\rm don} / M_{\rm acc} = 5.0$ 
($3.5$ in the case of the $M_{\rm don} = 16$\Msun model), and the
initial orbital periods were chosen such that the mass-loosing donors are consistent with the location of the \objname progenitor in the diagram, albeit for different mass transfer rates.}
\label{fig:HRD_binary}
\end{figure} 

Significant mass loss affects the luminosity of a donor star: instead of being radiated away from the surface, the 
energy from nuclear burning is captured in the expanding envelope layers and used to rearrange the internal structure \citep[e.g.,][]{Paczynski1967_MT}.
As luminosity is the key parameter used to estimate the mass of a star based on stellar tracks, it is essential to take this effect into account. This can only be achieved with detailed stellar-evolution computations.

Here, we employed the MESA stellar evolution code \citep{Paxton2011,Paxton2015,Paxton2019}\footnote{ 
MESA version r11554,  \url{http://mesa.sourceforge.net/}} to constrain binary models 
consistent with the \objname progenitor. The physical ingredients of our models 
(e.g., internal mixing, wind mass loss, chemical composition) 
are the same as in the reference model of \citet{Klencki2020a}, see their Sect.~2.
As previously discussed in $\S$ \ref{sec:single_stellar_models}, we adopt a metallicity of $Z = 0.0068$. Notably, we assume convective core overshooting
with the overshooting length of 0.345 pressure scale heights, as 
guided by observations of B-type giants ($\sim$15\,\Msun) in the LMC \citep{Brott2011}. 
This is roughly two times larger than the overshooting adopted in the MIST stellar tracks \citep[][calibrated to low-mass stars]{MIST2016ApJ},
and leads to a somewhat higher luminosity-to-mass ratio of our models. 

For simplicity, when computing binary models we only evolve the primary (donor) star and treat the companion as 
a point mass. The primary is initially nonrotating and can be spun-up by tidal interactions. 
We follow the formalism of \citet{Kolb1990} to calculate the mass-transfer rate through the L1 
Lagrangian point. We assume that the companion is unable to accrete any of the transferred mass.
This assumption is commonly made for systems evolving through a phase of rapid case B mass transfer based on a spin-up argument
\citep[once the accretor is quickly spun up, the accretion is expected to cease, e.g.,][]{deMink2013}. We note, however, that recent indirect evidence from Be X-ray binaries in the Small Magellanic Cloud suggests
a possibly higher accretion efficiency of $\sim$0.5 in such systems \citep{Vinciguerra2020}.
We further assume that specific angular momentum of the mass ejected from the system is an average 
of the angular momenta of the accretor and the L2 point, guided by the gas-kinematics study of \citet{MacLeod2020_preCE_massloss}.
This assumption is somewhat degenerate with the (unknown) mass ratio of the system: 
a fully isotropic mode of mass ejection (i.e., specific angular momentum of the accretor) would yield 
similar results to our models but for steeper mass ratios. 
We explore initial donor masses between 12 and 18 \Msun, mass ratios between 3 and 10, and orbital periods between 100 and 1000 days. 
The MESA inlists (input files) necessary to reconstruct our work
as well as all the binary models are available at \url{https://zenodo.org/communities/mesa}.  

In Fig.~\ref{fig:HRD_binary}, we show several tracks of donor stars in the HR diagram 
derived from binary models that were found consistent with the location of the \objname progenitor. 
In each case, as soon as the mass transfer rate increases above $\gtrsim 10^{-3} \rm M_{\rm \odot} \, yr^{-1}$, the donor's luminosity begins to decrease significantly. 
As a result, donors with various initial masses (ranging from 12 to 16 \Msun in Fig.~\ref{fig:HRD_binary}) 
can all be consistent with the locus of the \objname progenitor, albeit at different mass-transfer rates, ranging from $\log(\dot{M} / M_{\odot} \, {\rm yr}^{-1}) \approx -2.4$ for the 12 \Msun progenitor to $\approx -1.2$ for the 15 \Msun progenitor. The ages of the progenitors range from $\approx 12$\,Myr for the 16\,\Msun donor to $\approx 18.5$\,Myr for the 12\,\Msun donor.
The initial mass ratio was $q = 5$ for models with $M_{\rm don} = $ 12, 13, 14, and 15 \Msun, and $q = 3.5$
for the model with $M_{\rm don} = 16$\,\Msun.
Models with $M_{\rm don} = $ 17 or 18 \Msun were found inconsistent 
with the progenitor.
We note that models with $M_{\rm don} = 12-14$\,\Msun could also be consistent with the progenitor for all the other mass ratios explored here ($q$ from  $3.5$ to $10$). For a given donor mass, steeper mass ratios lead to higher mass transfer rates at the progenitor location ( see Sec.~\ref{sec:disc_binary}).

Fig.~\ref{fig:HRD_binary} is a clear illustration why the usage of binary rather than single stellar models is necessary to infer the initial progenitor mass. 
In Sect.~\ref{sec:disc_binary} we further discuss the interpretation of the \objname progenitor in view of the above analysis.

\subsection{Effect of binary mass loss on the progenitor appearance}
\label{sec:sph}

Many LRNe show a gradual brightening, lasting hundreds of days to several years before the main outburst \citep{Tylenda2011,Blagorodnova2017ApJ,Blagorodnova2020MNRAS,Pastorello2020_2}, which has been interpreted as an evidence of internal shocks in material leaving the binary system \citep{Pejcha2016,PejchaMetzger2017ApJ}. As a result, it is not immediately clear whether the progenitor detection corresponds to the underlying binary or the material leaving the binary. To elucidate the situation for \objname, we modified the smoothed-particle hydrodynamics code with radiative diffusion and cooling originally used by \citet{PejchaMetzger2017ApJ} to model the $\sim$200 day long pre-peak brightening in V1309~Sco. We set up several realizations of a binary system with component masses of $13$ and $2.6\,M_\odot$ on a circular orbit with separation of 1\,AU. We begin injecting mass around the outer Lagrange point with a rate of $3\times 10^{-2}\,M_\odot\,\text{yr}^{-1}$ (based on the analysis in Sect.~\ref{sec:binary_mesa_models}), which we increase as a power law toward a singularity at $2000$\,days from the start of the simulation. Within $1600$--$1800$\,days from the start of the simulation, the semi-major axis is reduced by a factor of $\sim$30 as the ejecta take away angular momentum of the binary, and the simulation stops due to limits on the equation of state. The luminosity of the outflow is initially around $2\times 10^3\,L_\odot$, which increases to about $10^4\,L_\odot$ near the end of the simulations. At this point, the mass-loss rate is a few solar masses per year. We conclude that the outflow luminosity $\sim$10\,years before the merger is much smaller than the luminosity of the stars, and the quantities derived in this paper represent the progenitor star.

\section{Explosion energetics}

Our photometric and spectroscopic analysis allow us to compute the main observables related to the photosphere during plateau. We estimate the ejecta velocity to be $v_{\rm{ej}}=500\pm65$\,\kms, as a weighted average of the FWHM of the \halpha line. The plateau has an average luminosity of $L_\text{p} = (1 \pm 0.15)\times 10^{40}$\,erg\,s$^{-1}$ and luminosity duration $t_\text{p}=70\pm 5$\,days (note that this duration is longer than in the optical bands). It is possible to invert the transient observables to estimate the physical parameters of the explosion. In this section, we take advantage of the binary-evolution models and confront the predictions of common-envelope energy formalism (Sect.~\ref{sec:ce_prescription}) with physical parameters inferred from the transient, interpreted either as a scaled-down version of Type II-P supernovae (Sect.~\ref{sec:recom}) or a shock-powered event (Sect.~\ref{sec:shock}).

\subsection{CE prescription}\label{sec:ce_prescription}

Classically, the outcome of a CE phase can be estimated based on the energy  formalism, in which the binding energy of the envelope, $E_{\rm bind}$, is related to the energy input from orbital inspiral, $\Delta E_{\rm orb}$, through a parameter $\alpha_{\rm CE}$ that is between 0 and 1: $E_{\rm bind} = \alpha_{\rm CE} \Delta E_{\rm orb}$ \citep{vdHeuvel1976,Tutukov1979IAUS,Webbink1984,Ivanova2013}. In our current understanding of CE evolution, events initiated by massive radiative-envelope donors (such as \objname) are generally expected to lead to stellar mergers rather than successful envelope ejections \citep{Klencki2020b}\footnote{Although the situation is less clear in the case of stellar accretors.}. For \objname, looking at the binary progenitor models with 12$-$16\,\Msun donors shown in Fig.~\ref{fig:HRD_binary} at the point when they are consistent with the observations, and the CE inspiral is about to commence, we find envelope binding energies of donors in the range $E_{\rm bind} \approx 2.5-3 \times 10^{49}\,\rm erg$ and energy inputs $\Delta E_{\rm orb} \approx 0.8 - 2 \times 10^{49}\,\rm erg$, indicating a merger even for an extreme assumption of $\alpha_{\rm CE} = 1.0$. For this estimate, $E_{\rm bind}$ includes a gravitational binding term, lowered by the internal energy of gas and radiation as well as the recombination energy.

\begin{figure}[h]
\includegraphics[width=0.5\textwidth]{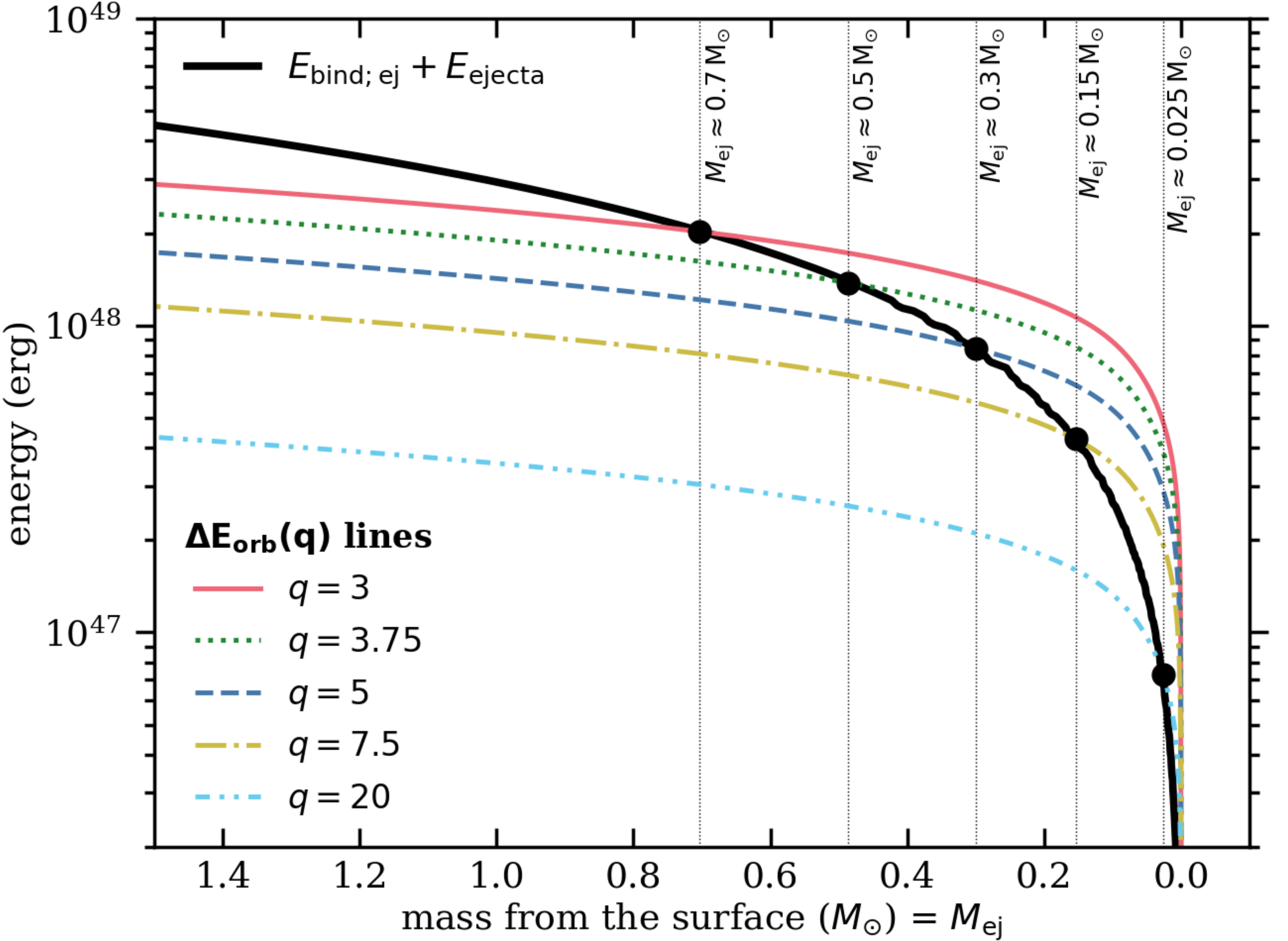} 
\caption{Estimation of the amount of mass ejected from the envelope during the \objname LRN event based on a simple energy formalism, (see Eqn.~\ref{eq:ce_presc_Mej}$-$\ref{eq:ce_presc_Mej4}). The envelope structure was taken from a binary MESA model with a 15 \Msun donor shown in Fig.~\ref{fig:HRD_binary}. The black line shows the energy required to unbind and bring to infinity a portion of the outer envelope with mass $M_{\rm ej}$. Coloured lines correspond to the orbital energy transferred by accretors with different masses when inspiraling deeper into the primary's envelope, up to the radius when the outer shell has mass $M_{\rm ej}$.  }
\label{fig:alpha_lambda_Mej}
\end{figure}

\begin{figure}
\includegraphics[width=0.5\textwidth]{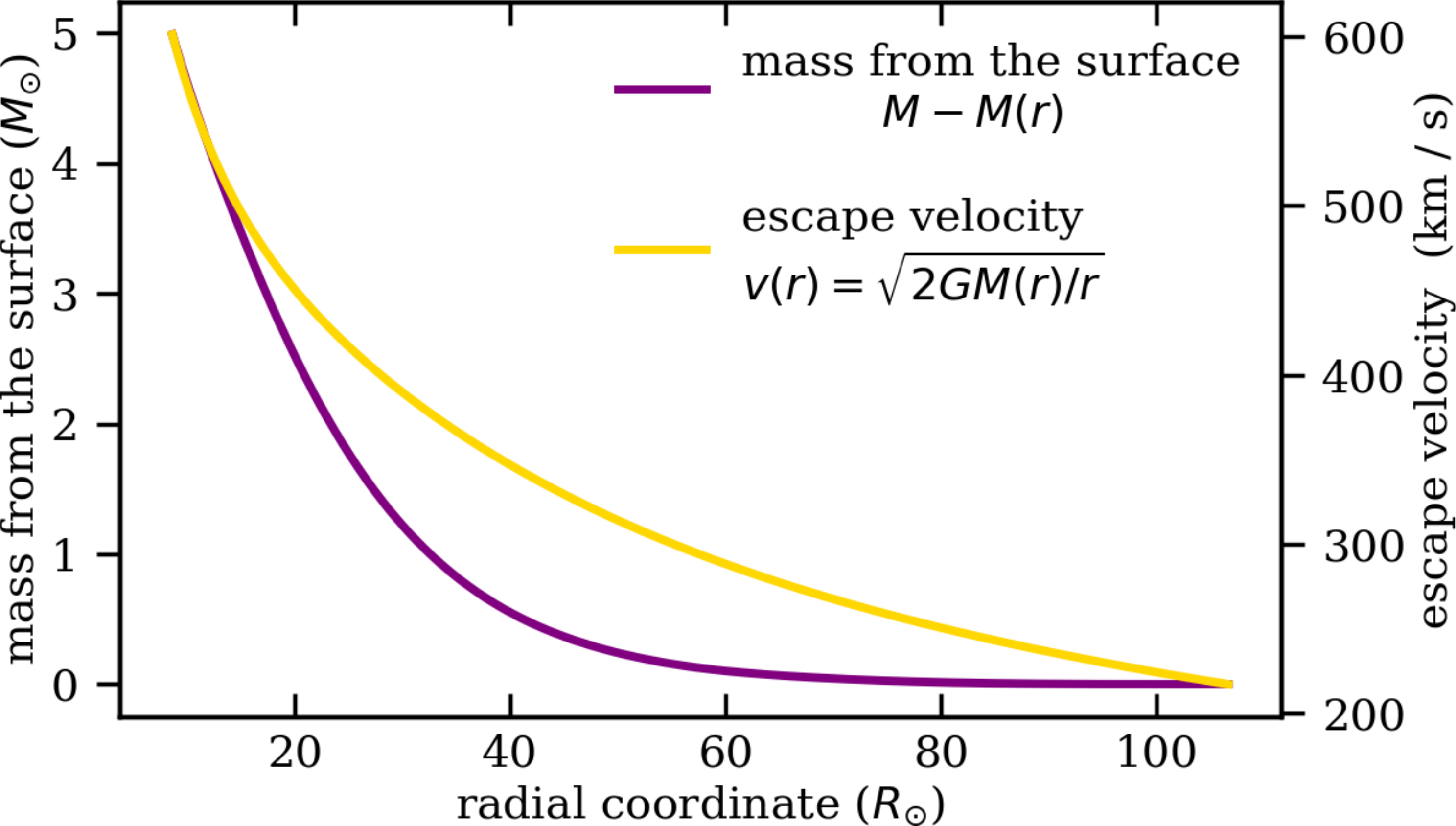} 
\caption{Escape velocity and mass from the surface as a function of the radial coordinate in the envelope of the progenitor of \objname derived from a binary evolution sequence of a 15 \Msun donor (initial mass) and a 3 \Msun accretor (see Sect.~\ref{sec:binary_mesa_models}). The selected envelope profile corresponds to when the donor star was consistent with the location of the \objname progenitor in the HR diagram, at which point its total mass was $\sim$ 13.1 \Msun, radius $\sim$ 107 \Rsun, and the mass transfer rate $\approx 6 \times 10^{-2} \rm M_{\rm \odot} \, yr^{-1}$ }.  
\label{fig:envelope}
\end{figure}

In the case of a merger, part of the loosely bound outer envelope becomes ejected, powering the LRN transient. Here, we estimate the mass of the ejecta, $M_{\rm ej}$, based on simple energy-budget considerations. We assume that: 

\begin{equation}
    E_{\rm bind;ej} + E_{\rm ejecta} = \Delta E_{\rm orb;Macc}
\label{eq:ce_presc_Mej}
\end{equation}
where the following terms are:
\begin{equation}
    E_{\rm bind;ej} = - \int_{M_{\rm don} - M_{\rm ej}}^{M_{\rm don}} \Bigg( - \frac{G M(r)}{r} + u \Bigg) \,{\rm d}m
    \label{eq:ce_presc_Mej2}
\end{equation}
\begin{equation}
E_{\rm ejecta} \approx 0.5 v_{\rm ej}^2 M_{\rm ej} - G (M_{\rm don}-M_{\rm ej}) M_{\rm ej}{R_{\rm ej}}^{-1}
\label{eq:ce_presc_Me3}
\end{equation}
\begin{equation}
\Delta E_{\rm orb;Macc} = -\frac{G M_{\rm don} M_{\rm acc}} {2 R_{\rm don}} + \frac{G (M_{\rm don}-M_{\rm ej}) M_{\rm acc}} {2 R_{\rm D}}.
\label{eq:ce_presc_Mej4}
\end{equation}

Here $E_{\rm bind;ej}$ is the binding energy of the ejected part of the envelope, $M_{\rm don}$ is the donor mass, $u$ is the internal energy per unit of mass, $E_{\rm ejecta}$ is the energy of the ejecta at infinity, $v_{\rm ej}$ is the velocity of the ejecta at radius $R_{\rm ej}$, and $E_{\rm orb;Macc}$ is the energy input from inspiral of the accretor with mass $M_{\rm acc}$ starting from the donor surface $R_{\rm don}$ down to some radius $R_{\rm D}$ at the bottom of the ejected part of the envelope (corresponding to $M_{\rm ej}$). We assume $v_{\rm{ej}}=500\pm65$\,\kms and $R_{\rm ej} \approx R_{\rm don} \approx$ 110 \Rsun. The exact value of $R_{\rm ej}$ does not influence the result, as the kinetic term dominates the ejecta energy.

In Fig.~\ref{fig:alpha_lambda_Mej} we solve Eq.~\ref{eq:ce_presc_Mej} for the envelope structure taken from the binary model with a 15 \Msun donor (initial mass) shown in the HR diagram in Fig.~\ref{fig:HRD_binary}. The result is plotted as a function of the mass from the surface of the envelope, i.e., the ejected amount $M_{\rm ej}$. Coloured lines assume different initial mass ratios $q$ ranging from 3 to 20, and black dots indicate solutions to Eq.~\ref{eq:ce_presc_Mej}. For the specific case of a $15\,M_\odot$ progenitor, we also show in Fig.~\ref{fig:envelope} the envelope structure of mass, radius, and local escape velocity. We see that the mass from the surface starts to dramatically increase only when we get closer to the core and that the local escape velocity increases more gradually.

The case with $q = 20$, possibly related to a dynamical CE onset (see discussion in Sect.~\ref{sec:disc_binary}), ejects only $\approx 0.025$ \Msun of material. This amount appears to be an order of magnitude too low to explain the LRN transient energetics, as further discussed in Sects.~\ref{sec:recom} and \ref{sec:shock}.

The case with $q = 3$ did not lead to an unstable mass transfer and a CE evolution in our binary MESA simulations (minding the caveat that we model the accretor as a point mass) and is shown in Fig.~\ref{fig:alpha_lambda_Mej} as a limiting case.  Examples with $q$ between 3.75 and 7.5 match well the binary evolution scenario for \objname, in which a phase of thermal-timescale mass transfer is culminated in a dynamical inspiral as modeled in Sect.~\ref{sec:binary_mesa_models}. In these cases, the energy-budget estimation suggests ejecta masses of $M_{\rm ej} \approx 0.15-0.5$ \Msun.

We carried out this exercise for all the donor masses that were found consistent with the progenitor in our binary model analysis (see Fig.~\ref{fig:HRD_binary}), finding a very similar range $M_{\rm ej} \approx 0.15-0.5$\,\Msun for donors with masses 12, 13, and 14\,\Msun. The case with $M_{\rm don} = 16$\,\Msun yielded a higher value $M_{\rm ej} \approx 0.8-1.0$\,\Msun; the difference is primarily the result of a larger fraction of the donor mass being stripped in the mass transfer prior to the merger in the 16\,\Msun case (see also Fig.~\ref{fig:Mdot_time} in the discussion).

We note that between the progenitor observations (10$-$13\,yr before the merger) and the LRN event itself, the donor is expected to still lose a substantial amount of mass of $\gtrsim 1$\,\Msun. Therefore, the envelope structure models used here are only an approximation of the envelope structure at the onset of the dynamical inspiral. As such, the inferred values of $M_{\rm ej}$ should be treated as a crude estimation, mainly indicating an order of magnitude for $M_{\rm ej}$.
In the following sections we constrain the ejected mass in more detail based on models of LRNe energetics.

\subsection{Scaled Type II-P supernova model }
\label{sec:recom}

Light curves of hydrogen-rich explosive transients will generally show plateaus of nearly constant luminosity, as the photosphere is located near the hydrogen recombination front, where the opacity steeply changes over a small range of temperatures. This is commonly seen in Type II-P supernovae, but a similar model was developed by \citet{Ivanova2013} to link the observational class of LRNe with common-envelope ejection events. 

Both Type II-P supernovae and LRNe have similar plateau durations of $\sim$100\,days, but Type II-P supernovae have ejecta masses $\sim$ 20 times higher and expansion velocities $\sim$10 times higher than LRNe. As a result, LRNe ejecta are a factor of $\sim$50 denser than those of Type II-P supernovae. Still, we expect the equation of state of optically thick LRN ejecta to be radiation-dominated with density-independent Thomson scattering as the primary opacity source. Around or below hydrogen recombination, the opacity might depend on density. For example, if the Rosseland-mean opacity of the negative hydrogen ion, which scales as $\rho^{0.5}$, dominates in the outer LRN ejecta, we expect that effective temperatures of LRNe will be lower than those of Type II-P supernovae. However, the plateau duration or luminosity should not strongly depend on the form of low-temperature opacity. The only fundamental difference between Type II-P supernovae and LRNe could be the density profile of the ejecta, which depends on the uncertain mass-ejection mechanism in LRNe.

Given these similarities between Type II-P supernovae and LRNe, we follow \citet{Ivanova2013} and apply the same analytic scaling relations linking plateau luminosity $L_\text{p}$, duration $t_\text{p}$, and expansion velocity $v_\text{ej}$ to the explosion energy $E$, ejecta mass $M_\text{ej}$, and initial radius $R_0$ \citep{arnett80,litvinova85}. We explore different calibrations for these relations from the literature.

Recently, \citet{sukhbold16} found an excellent agreement between 1D flux-limited diffusion simulations of Type II-P supernova light curves and analytic scaling relations of \citet{popov93} with modified absolute coefficients. The plateau luminosity $L_\text{p}$ is
\begin{equation}
    L_\text{p} = 9.24 \times 10^{38}\,\text{erg\,s}^{-1} \left( \frac{M_\text{ej}}{M_\odot}\right)^{1/3} \left(\frac{v_\text{ej}}{100\,\text{km\,s}^{-1}}  \right)^{5/3} \left(\frac{R_0}{100\,R_\odot}   \right)^{2/3},
    \label{eq:s16_l}
\end{equation}
and the plateau duration $t_\text{p}$
\begin{equation}
    t_\text{p} = 108\,\text{days} \left( \frac{M_\text{ej}}{M_\odot}\right)^{1/3} \left(\frac{v_\text{ej}}{100\,\text{km\,s}^{-1}}  \right)^{-1/3} \left(\frac{R_0}{100\,R_\odot}   \right)^{1/6},
    \label{eq:s16_t}
\end{equation}
where we replaced $E = \frac{1}{2}M_\text{ej} v_\text{ej}^2$ and rescaled to parameters relevant for LRNe.

\citet{Ivanova2013} and \citet{MacLeod2017ApJ} used identical equations but with the absolute term based on the original work of \citet{popov93}. In our units, their absolute terms are $4.2 \times 10^{38}\,\text{erg s}^{-1}$ and $133$\,days for their default choice of opacity of $0.32\,\text{cm}^2\,\text{g}^{-1}$ and recombination temperature of $4500$\,K. Alternatively, \citet{kasen09} performed Monte Carlo multiwavelength radiation transport in the context of Type II-P supernovae and obtained an absolute term in  Eq.~(\ref{eq:s16_l}) of $6.29 \times 10^{38}\,\text{erg\,s}^{-1}$ and slightly different exponents in Eq.~(\ref{eq:s16_t}), which then becomes
\begin{equation}
    t_\text{p} = 295\,\text{days} \left( \frac{M_\text{ej}}{M_\odot}\right)^{1/4} \left(\frac{v_\text{ej}}{100\,\text{km\,s}^{-1}}  \right)^{-1/2} \left(\frac{R_0}{100\,R_\odot}   \right)^{1/6}.
    \label{eq:k09_t}
\end{equation}
However, the calibration of \citet{kasen09} might be less relevant for LRNe, because their simulations are tuned for Type II-P supernova densities and velocities. In Eqs.~(\ref{eq:s16_l}--\ref{eq:k09_t}), $R_0$ should be interpreted as a radius where any heating of the ejecta stops, and the ejecta expands homologously thereafter \citep{goldberg20}. 

To find $M_\text{ej}$ and $R_0$ from Eqs.~(\ref{eq:s16_l}--\ref{eq:k09_t}), we randomly sampled $L_\text{p}$, $t_\text{p}$, and $v_\text{ej}$ assuming uncorrelated Gaussian distributions. For the \citet{sukhbold16} calibration, we find $M_\text{ej} = 2.03^{+0.46}_{-0.39}\,M_\odot$ and $R_0 = 44^{+13}_{-10}\,R_\odot$, while for the coefficients of \citet{MacLeod2017ApJ} we find $M_\text{ej} = 0.40^{+0.09}_{-0.08}\,M_\odot$ and $R_0 = 326^{+97}_{-72}\,R_\odot$. For the \citet{kasen09} calibration we find $M_\text{ej} = 0.019^{+0.008}_{-0.006}\,M_\odot$ and $R_0 = 809^{+315}_{-223}\,R_\odot$. All values stated here are medians of the distributions, and the confidence intervals include 68\% of the realizations. In all cases, the inferences of $M_\text{ej}$ and $R_0$ are highly anti-correlated in the sense that higher values of $M_\text{ej}$ are accompanied by lower values of $R_0$. The correlation coefficients between $\log M_\text{ej}$ and $\log R_0$ are around $-0.95$. The observed velocity of about $500\,\text{km s}^{-1}$ implies that hydrogen recombination was subdominant in powering the ejection, because the velocity scale of hydrogen recombination is only about $50\,\text{km s}^{-1}$.

It is interesting to note the substantial differences in the inferred values from the relations of \citet{sukhbold16} and \citet{MacLeod2017ApJ}, which differ only by the absolute terms.  The estimate of $M_\text{ej}$ based on \citet{sukhbold16} is substantially larger than predicted in Sect.~\ref{sec:ce_prescription} (Fig.~\ref{fig:alpha_lambda_Mej}), but $R_0$ is much smaller and located inside the progenitor. We see that $M_\text{ej}$ and $R_0$ are roughly compatible with the progenitor structure and that $v_\text{ej}$ corresponds to the escape velocity at $R_0$ within few tens of percent, as we showed in Sect.~\ref{sec:ce_prescription} (Fig.~\ref{fig:envelope}). It is natural to expect that the common-envelope ejecta velocity will be similar to the escape velocity of the binary companion at radius $r$, where most of the envelope ejection occurs. 

The $M_\text{ej}$ based on \citet{MacLeod2017ApJ} is roughly compatible with the predictions of the CE theory shown in Sect.~\ref{sec:ce_prescription} (Fig.~\ref{fig:alpha_lambda_Mej}), however, $R_0$ is substantially larger than the initial progenitor radius, implying that the ejecta got reheated after ejection, probably by internal shocks.

The inferences based on \citet{kasen09} relations give very small $M_\text{ej}$ and unrealistically large $R_0$, much larger than the original progenitor radius or the likely binary orbit. The likely explanation is that the \citet{kasen09} relations are too tuned for Type II-P supernovae and hence not applicable to LRNe.

To summarize our results, the inferences based on recombination models are broadly compatible with other constraints on the progenitor if the event was a stellar merger, although there are discrepancies with the standard common-envelope energy formalism applied to binary evolution models of the \objname progenitor. Analytic scaling relations give different values of $M_\text{ej}$ and $R_0$. Our preferred choice of analytic scaling relation is the one of \citet{sukhbold16}. Analytic scaling relations are especially powerful for examining relative differences in a population of objects, where the uncertain absolute terms play less of a role.

\subsection{Shock-powered model}
\label{sec:shock}

\citet{MetzgerPejcha2017MNRAS} proposed a model for the light curves of LRNe, in which faster spherically symmetric ejecta collide with a preexisting equatorially concentrated mass distribution. This model can naturally explain double-peaked LRN light curves, where the first peak is caused by cooling of the fast unshocked polar ejecta, while the second peak or plateau arises due to continuing shock interaction acting as an embedded power source inside the ejecta. The equatorial distribution is formed by mass loss from the binary preceding the dynamical merger, which is likely caused by nonconservative mass transfer and/or by ejection through the outer Lagrange point \citep{Shu1979ApJ,Pejcha2014ApJ}. This long-lasting pre-dynamical mass loss can explain gradual brightenings seen in many LRNe before the primary peak, as was quantified for V1309~Sco by \citet{PejchaMetzger2017ApJ}.

A power source embedded inside hydrogen-rich ejecta will in many cases lead to a plateau-like light curve. \citet{sukhbold17} provided a particularly striking example by embedding a spinning-down magnetar inside otherwise sub-luminous Type II-P supernovae ejecta. This setup led to plateau luminosities higher by a factor of $\sim$30 and light curves photometrically indistinguishable from normal Type II-P supernovae. Another possible source of internal heating could also be jets emanating from a central engine \citep{Shiber2019MNRAS,Soker2020ApJ,SokerKaplan2021}.

Within the model of spherical ejecta colliding with a preexisting equatorial mass distribution, \citet{MetzgerPejcha2017MNRAS} provided analytic relations for the characteristic luminosity of the shock-powered peak
\begin{equation}
    L_\text{p} = \frac{9}{32 f_\Omega} \frac{v_\text{sh}^3}{v_\text{w}}\frac{M_\text{w}}{t_\text{run}} \exp\left(-\frac{v_\text{sh}t_\text{p}}{v_\text{w} t_\text{run}} \right),
    \label{eq:shock_l}
\end{equation}
and the timescale for the shock-powered light curve to rise to its peak
\begin{equation}
    t_\text{p} = \left(\frac{M_\text{ej}\kappa}{6 \pi c v_\text{sh}}   \right)^{1/2},
    \label{eq:shock_t}
\end{equation}
where we identify $L_\text{p}$ with the plateau luminosity and $t_\text{p}$ with the plateau duration. Here, $M_\text{w}$ is the mass of the preexisting wind, which occupies a fraction of the solid angle $f_\Omega$ and which was formed by binary mass loss with a rate exponentially increasing on a timescale $t_\text{run}$. The radial velocity of the preexisting wind is $v_w$, which we assume to be 10\% of the fast spherically symmetric ejecta velocity $v_\text{ej}$ with mass $M_\text{ej}$. Following \citet{MetzgerPejcha2017MNRAS}, we assume that the velocity of the shocked shell is $v_\text{sh} = 0.5 v_\text{ej}$. We further assume opacity $\kappa = 0.32\,\text{cm}^2\,\text{g}^{-1}$.

Following the same procedure as in Sect.~\ref{sec:recom}, we can use the observed properties of \objname to infer $M_\text{w}$ and $M_\text{ej}$ from Eqs.~(\ref{eq:shock_l}--\ref{eq:shock_t}). For $t_\text{run} = 1000$\,days, we get $M_\text{w} = 2.34^{+0.31}_{-0.27}\,M_\odot$ and $M_\text{ej} = 0.80^{+0.07}_{-0.06}\,M_\odot$. 

To our knowledge, this is the first time that Eqs.~(\ref{eq:shock_l}--\ref{eq:shock_t}) have been used to infer parameters of a transient, which makes it worthwhile to explore the sensitivity to changes of parameters. Increasing or decreasing $t_\text{run}$ by a factor of 10 leads to $M_\text{w}$ higher by a factor of few. Fixing $v_\text{w}$ to $100$ or $10$\,km\,s$^{-1}$ changes $M_\text{w}$ to $1.92^{+0.21}_{-0.19}$ or $3.92^{+0.70}_{-0.57}\,M_\odot$, respectively. If we put a pre-factor of $2$ or $0.5$ in front of the brackets of Eq.~(\ref{eq:shock_t}), which represents our uncertainty in relating the observed transient duration to the theoretical timescale, $M_\text{ej}$ changes to $0.20^{+0.02}_{-0.02}$ or $3.22^{+0.26}_{-0.25}\,M_\odot$, respectively. In all cases, the correlation coefficient between $\log M_\text{ej}$ and $\log M_\text{w}$ is around $-0.52$, which does not suggest any strong degeneracy between the two parameters.

Although the analytic model of \citet{MetzgerPejcha2017MNRAS} has not yet been calibrated with a multidimensional radiation hydrodynamic simulation, our inferred parameter values appear reasonable. We find that the amount of mass lost in the pre-dynamical phase of the merger is likely a few solar masses. Binary-evolution models predict that the mass-transfer rate steeply increases with time (Fig.~\ref{fig:Mdot_time}), and we expect that most of the material leaves the binary. Our simulations of the final few thousand days of the binary evolution described in Sect.~\ref{sec:sph} suggest that about $2\,M_\odot$ of material lost with specific angular momentum of the outer Lagrange point is sufficient to remove most of the binary angular momentum. Our inference of $M_\text{w}$ is thus compatible with the theoretical expectations. The value of $M_\text{ej}$ also agrees with the predictions of common-envelope energy formalism shown in Fig.~\ref{fig:alpha_lambda_Mej}, especially if the plateau duration is longer than indicated by Eq.~(\ref{eq:shock_t}). There is the same discrepancy as in the recombination model that the escape velocity of the progenitor layer corresponding to $M_\text{ej}$ is by a factor of two smaller than the observed $v_\text{ej}$.

To summarize, both recombination-only and shock-interaction models can explain the observed properties of \objname within the limits of current theoretical uncertainties and degeneracies inherent to the inverse problem. A more involved comparison would require developing the shock-interaction model to the same level of sophistication as the recombination model achieved for Type II-P supernovae. Parameters from both models are broadly compatible with the common-envelope  theory (for mass ratios $q \lesssim 10$) and the progenitor structure of \objname. Further study should be devoted to the discrepancy between the observed $v_\text{ej}$, which implies mass ejection from deep inside the progenitor, and the relatively low $M_\text{ej}$, which suggests ejection of layers close to the surface.

\section{Discussion}

This is the first observational study of an extragalactic stellar merger in which the progenitor was interpreted using binary stellar-evolution models. In this section, we first discuss \objname in the context of other LRNe. Next, we elaborate on the interpretation, limitations, and future directions for modelling LRN progenitors using stellar binary models. Finally, we consider the binary-evolution model within the context of the star cluster where \objname is located.

\subsection{Comparison of \objname with other LRNe}

\begin{figure*}[h]
\centering
\includegraphics[width=\linewidth]{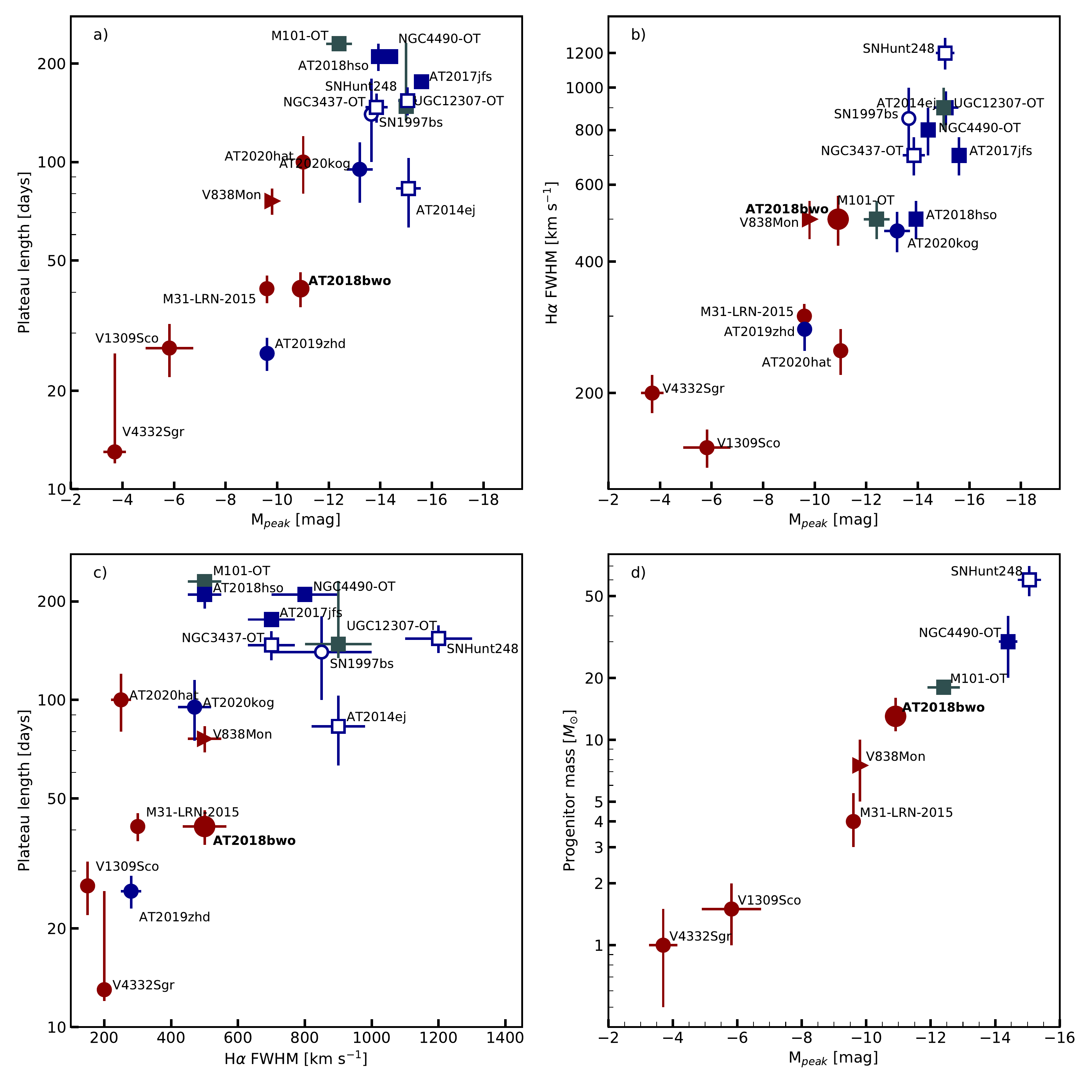}
\caption{Top left: Peak absolute magnitude in $V$, $R$ or $r$-band versus the length of the plateau in that band. Top right: Peak absolute magnitude vs. the FWHM of the \halpha line, which is a proxy for the outflow expansion velocity. Bottom left: Plateau length vs. \halpha FWHM. Bottom right: Estimated mass of the progenitor star vs. transient peak magnitude in $V$, $R$ or $r$-bands. Filled circles represent LRNe that developed molecular absorption bands at later times. Open circles indicate additional LRNe candidates whose nature is still debated. Circles represent LRNe with one initial peak followed by a plateau. Squares are LRNe having a slower secondary red peak. Triangles show LRNe with 3 maxima. Red markers represents transients that already showed red continuum during the first peak. Blue markers correspond to LRNe that showed a hotter continuum and Balmer emission lines at early times. Gray markers are used for transients without early times spectroscopy. The name of some transients has been shortened for clarity.}
\label{fig:comparison}
\end{figure*}

\begin{table*}
\renewcommand{\tabcolsep}{0.11cm}
\centering
\begin{tabular}{ccccccccc}
Object & Distance & $E(B-V)$ & Filter & M$_{\rm{peak}}$ & t$_{\rm{plateau}}$  & FWHM     & Progenitor mass & References  \\ 
       &    (pc) &	(mag)  &      &	(mag)      & (days)              &  (\kms)   &   (\Msun) & \\ \hline\hline
V1309Sco & (2.10$\pm$0.70)$\times10^{3}$ & 0.550 & $V$ & -5.82$\pm$0.92 & 27$\pm$5 & 150$\pm$15 & 1.5$_{-0.5}^{+0.5}$ & [1] [2] [3]  \\
V4332Sgr & (5.00$\pm$1.00)$\times10^{3}$ & 0.320 & $V$ & -3.70$\pm$0.44 & 13$\pm$10 & 200$\pm$20 & 1.0$_{-0.5}^{+0.5}$ & [4] [3]  \\
V838Mon & (5.90$\pm$0.40)$\times10^{3}$ & 0.850 & $R$ & -9.80$\pm$0.15 & 76$\pm$7 & 500$\pm$50 & 7.5$_{-2.5}^{+2.5}$ & [5] [6] [7] [8]  \\
M31-LRN2015 & (0.76$\pm$0.02)$\times10^{6}$ & 0.120 & $R$ & -9.60$\pm$0.20 & 41$\pm$4 & 300$\pm$20 & 4.0$_{-1.0}^{+1.5}$ & [9] [10] [11]  \\
AT2019zhd & (0.76$\pm$0.02)$\times10^{6}$ & 0.055 & $r$ & -9.61$\pm$0.08 & 26$\pm$3 & 280$\pm$30 & $-$ & [12]  \\
AT2020hat & (5.16$\pm$0.21)$\times10^{6}$ & 0.090 & $r$ & -11.01$\pm$0.10 & 100$\pm$20 & 250$\pm$30 & $-$ & [13]  \\
M101-OT2015-1 & (6.43$\pm$0.20)$\times10^{6}$ & 0.008 & $R$ & -12.40$\pm$0.50 & 230$\pm$10 & 500$\pm$50 & 18.0$_{-1.0}^{+1.0}$ & [14]  \\
AT2018bwo & (6.64$\pm$0.10)$\times10^{6}$ & 0.029 & $r$ & -10.91$\pm$0.03 & 41$\pm$5 & 500$\pm$65 & 13.0$_{-2.0}^{+3.0}$ & [15]  \\
SN1997bs* & (9.20$\pm$0.60)$\times10^{6}$ & 0.210 & $R$ & -13.65$\pm$0.14 & 140$\pm$40 & 850$\pm$150 & $-$ & [16] [17]  \\
NGC4490-2011OT1 & (9.60$\pm$1.40)$\times10^{6}$ & 0.320 & $R$ & -14.40$\pm$0.29 & 210$\pm$10 & 800$\pm$100 & 30.0$_{-10.0}^{+10.0}$ & [18] [17]  \\
NGC3437-2011OT1* & (20.90$\pm$4.60)$\times10^{6}$ & 0.020 & $R$ & -13.85$\pm$0.43 & 147$\pm$15 & 700$\pm$70 & $-$ & [17]  \\
AT2018hso & (21.30$\pm$0.56)$\times10^{6}$ & 0.300 & $r$ & -13.93$\pm$0.06 & 210$\pm$20 & 500$\pm$50 & $-$ & [19]  \\
AT2014ej* & (22.10$\pm$1.50)$\times10^{6}$ & 0.310 & $r$ & -15.09$\pm$0.48 & 83$\pm$20 & 900$\pm$80 & $-$ & [20]  \\
AT2020kog & (22.50$\pm$2.00)$\times10^{6}$ & 0.370 & $r$ & -13.20$\pm$0.50 & 120$\pm$30 & 470$\pm$50 & $-$ & [13]  \\
SNHunt248* & (22.50$\pm$4.00)$\times10^{6}$ & 0.040 & $R$ & -15.06$\pm$0.36 & 154$\pm$15 & 1200$\pm$100 & 60.0$_{-10.0}^{+10.0}$ & [21] [22]  \\
AT2017jfs & (35.70$\pm$2.70)$\times10^{6}$ & 0.022 & $r$ & -15.60$\pm$0.17 & 176$\pm$5 & 700$\pm$70 & $-$ & [23]  \\
UGC12307-2013OT1 & (39.70$\pm$2.80)$\times10^{6}$ & 0.220 & $R$ & -15.00$\pm$0.15 & 148$\pm$72 & 900$\pm$100 & $-$ & [17]  \\
\hline \end{tabular} \caption{Sample of LRNe shown in Fig.~\ref{fig:comparison}. Magnitudes in $V$ and $R$ filters are reported in the Vega magnitude system. Magnitudes in the $r$ filter are reported in the AB system. The sign $^*$ indicates LRN candidates. References: [1]: \citet{Mason2010}, [2]: \citet{Tylenda2011}, [3]: \citet{Kaminski2018AA}, [4]: \citet{Martini1999}, [5]: \citet{Munari2002}, [6]: \citet{Ortiz-Leon2020}, [7]: \citet{Kimeswenger2002MNRAS}, [8]: \citet{Afcsar2007AJ}, [9]: \citet{Williams2015}, [10]: \citet{Kurtenkov2015AA}, [11]: \citet{MacLeod2017ApJ}, [12]: \citet{Pastorello2020_1}, [13]: \citet{Pastorello2020_2}, [14]: \citet{Blagorodnova2017ApJ}, [15]: This work, [16]: \citet{VanDyk2000PASP}, [17]: \citet{Pastorello2019b}, [18]: \citet{Smith2016b}, [19]: \citet{Cai2019AA}, [20]: \citet{Stritzinger2020AA}, [21]: \citet{Kankare2015}, [22]: \citet{Mauerhan2015}, [23]: \citet{Pastorello2019a},  \label{tab:lrn_sample} }
\end{table*}

Our LRNe sample contains known transients from the literature (see Table \ref{tab:lrn_sample}), except two Galactic LRNe OGLE-2002-BLG-360 \citep{Tylenda2013} and CK Vul \citep{Kato2003AA}, and the LRN in M31: M31\,RV \citep{Rich1989,Mould1990}, due to the limited available data. Our comparison, shown in Fig.~\ref{fig:comparison}, particularly focuses on the correlations between: a) the peak magnitude and the length of the plateau, b) the peak magnitude and the FWHM of the \halpha line reported in the literature, which is a proxy for the expansion velocity of the ejecta, c) \halpha FWHM and the length of the plateau, and d) the peak magnitude and the inferred mass for the primary progenitor. Here we distinguish between bona-fide LRNe that showed molecular absorption at later times and LRNe candidates, where such absorption was not identified (filled vs. empty circles). In addition, we marked those LRNe that at early times showed a cool, red continuum (red markers) from those that appeared hotter (blue markers) and showed stronger Balmer emission lines from interaction with the circumstellar medium (CSM).

 One possible source of confusion for LRNe is another ``gap transients'' class of intermediate luminosity red/optical transients (ILOT/ILRT), such as SN\,2008S \citep{Prieto2008,Botticella2009}, NGC\,300\,OT2008-1 \citep{Bond2009,Humphreys2011}, iPTF\,10fqs \citep{Kasliwal2011} or M85-OT (initially proposed to be a LRN by \citet{Kulkarni2007}). The progenitors of these transients are generally massive, dust-enshrouded stars \citep{Prieto2009ApJ}, and their spectra are characterized by generally cold continuum with emission for the Balmer and \io{Ca}{ii} IR triplet lines, as well as the characteristic lines of [\io{Ca}{ii}], which are absent in LRNe spectra. The colour evolution for ILRTs is also more moderate when compared to LRNe, and their late-time spectroscopy does not show the appearance of molecular absorption bands.
 
Within the LRNe sample, \objname corresponds to a transient with intermediate brightness and duration. Its peak magnitude of $M_r = -10.91 \pm 0.03$\,mag is in between V838\,Mon and M101\,OT2015-1, similar to AT\,2020hat. However, its $r$-band plateau duration of $\sim$41\,days is in better agreement with the less luminous M31-LRN2015. This shorter timescale may be caused by the higher ejecta velocity of $\sim$500\,\kms as compared to other lower luminosity LRNe, which all had FWHM $\leq$300\,\kms.

One interesting highlight of the comparison is the strong correlation between the peak magnitude of the transient and other observables, e.g., the length of the plateau, the expansion velocity, and the mass of the progenitor star. Similar correlations were explored in \cite{Kochanek2014,Mauerhan2018MNRAS,Pastorello2019a,Pastorello2020_2}, although with slightly different approach. For example, our analysis estimates the duration of the transient by fitting an analytical function to the optical light curve, as described in \cite{Valenti2016MNRAS} for core collapse supernovae (see their eq.~1). We considered the time of the first peak as our reference epoch and adopted an uncertainty of 10\%. For UGC\,12307-2013OT1 the time of the first peak is unknown, so that we computed the plateau duration for the available part of the lightcurve and included the time to the last nondetection as an upper error in our measurement. For V4332\,Sgr, there is no available data before its discovery, so we also cautiously adopted a larger upper value for the duration of the plateau.

Another highlight is the apparent distinction between LRNe depending on their peak magnitude. On the one hand, transients with $M_{\rm peak} \geq -12$\,mag generally show one or no distinguishable peaks, followed by a plateau in the optical bands. At early times, this group also mostly shows a reddened continuum and a forest of narrow absorption lines. On the other hand, brighter transients tend to show two distinct peaks: an initial fast ``blue'' peak, followed by a more extended ``'red'' peak. Spectroscopy taken during the first peak shows a rather hot continuum with strong Balmer emission lines from CSM interaction. During the second peak, the characteristic cool continuum emerges, and molecular lines form. If Thompson scattering is the main mechanism responsible for the formation of emission lines, analogous to interacting SNe, then the two groups would differ in the state of ionization of the photosphere during peak. 

For higher-luminosity events, the photosphere would initially be located in a fully ionized shell of previously ejected material. This shell, preceding the shock wave generated by the mass ejection during the merger, would reprocess the shock emission until the ejecta take over, pushing the rapidly cooling photosphere outwards. This cooling would also move the peak of the continuum emission toward longer wavelengths, appearing as a secondary peak in the redder bands. While during the first peak the radius is expected to remain fairly constant, the cooling and expansion of the ejecta would increase the apparent location of photosphere while keeping the overall luminosity relatively stable. 

For lower-luminosity events, the photosphere would be located in a neutral (or partially ionized) shell. Therefore, any shock emission generated by the dynamical ejection of mass would be absorbed and reprocessed by this layer, transforming the shock's kinetic energy into thermal energy that would heat and partially ionize the shell. Spectroscopically, the shell would still appear as a reddened continuum at early times, as the outer neutral layers would absorb great part of the emission at shorter wavelengths.

The ionization of the outer shell appears to depend on the peak magnitude of the outburst. Brighter events with potentially more massive progenitors are likely capable of generating more energetic outflows and supply enough radiation to fully ionize the shell before the onset of the dynamical ejection.

When comparing the length of the LRNe plateau and the expansion velocity (from \halpha FWHM), there seems to be yet another dichotomy between lower- and higher-luminosity events. While fainter and redder events have generally some correlation between \halpha FWHM and plateau length, the higher-luminosity group has a much larger spread. This is mainly due to transients we designated as LRNe candidates (empty markers), which seem to fall outside of the increasing trend, as they correspond to shorter events with larger expansion velocities.

The final highlight is the tight correlation between the mass of the primary progenitor and the peak magnitude of the outburst, even for LRNe with massive progenitors. This trend was first proposed by \citet{Kochanek2014} based on Galactic LRNe with masses lower than 10\,\Msun. Our sample shows that such a correlation can also be extended to extragalactic events, with generally more massive progenitors. This shows that the amount of mass ejected in stellar mergers is proportional to the initial mass of the donor, although fluctuations may be associated with the mass of the secondary companion or the geometry of the system. 

\subsection{Binary stellar models for LRN progenitors}
\label{sec:disc_binary}

In this work, we analyzed the progenitor of \objname in the rightful context of binary evolution. We interpret the star observed 10$-$13\,yr before the LRN event as 
a primary (more massive and luminous) component in a binary system in which a phase of mass transfer culminated in a common-envelope event and the LRN transient. 
In Sect.~\ref{sec:binary_mesa_models}, with the use of a grid of binary stellar models computed with MESA, we found that the location of the \objname progenitor in the HR diagram can be consistent with a donor star of an initial mass in the range $\sim$12$-$16\,\Msun (see Fig.~\ref{fig:HRD_binary}). This mass is significantly larger than inferred from single-star models ($\sim$11\,\Msun, see Sect.~\ref{sec:single_stellar_models}), which highlights the importance of binary models for LRN progenitors. \\

 \begin{figure}
{\includegraphics[width=0.5 \textwidth]{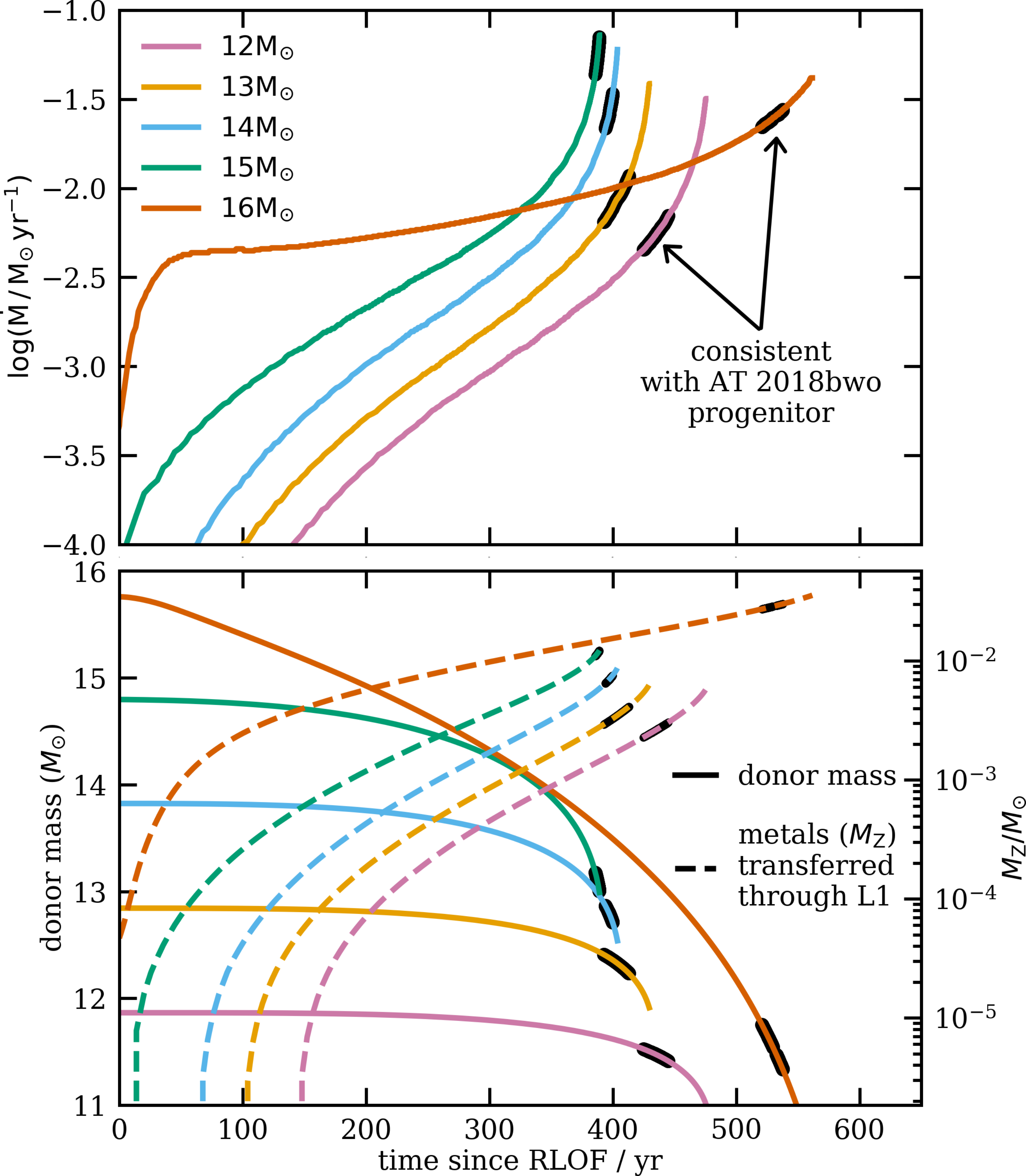} }
\caption{Time evolution of the mass transfer rate (upper panel) and the donor mass as well as the cumulative mass of metals $M_Z$ transferred through L1 since RLOF (lower panel) in five exemplary binary models in which the position of the donor in the HR diagram is consistent with the \objname progenitor (see Fig.~\ref{fig:HRD_binary}). The mass transfer rate at which the donor is consistent with the \objname progenitor (marked with bold in the figure) depends on the initial mass of the donor. The evolution of $\dot{M}$ is distinctively different in the 16  \Msun case because of a shallow outer convective zone in the envelope of the 16 \Msun donor at RLOF.}
\label{fig:Mdot_time}
\end{figure}

\begin{figure}
{\includegraphics[width=0.5 \textwidth]{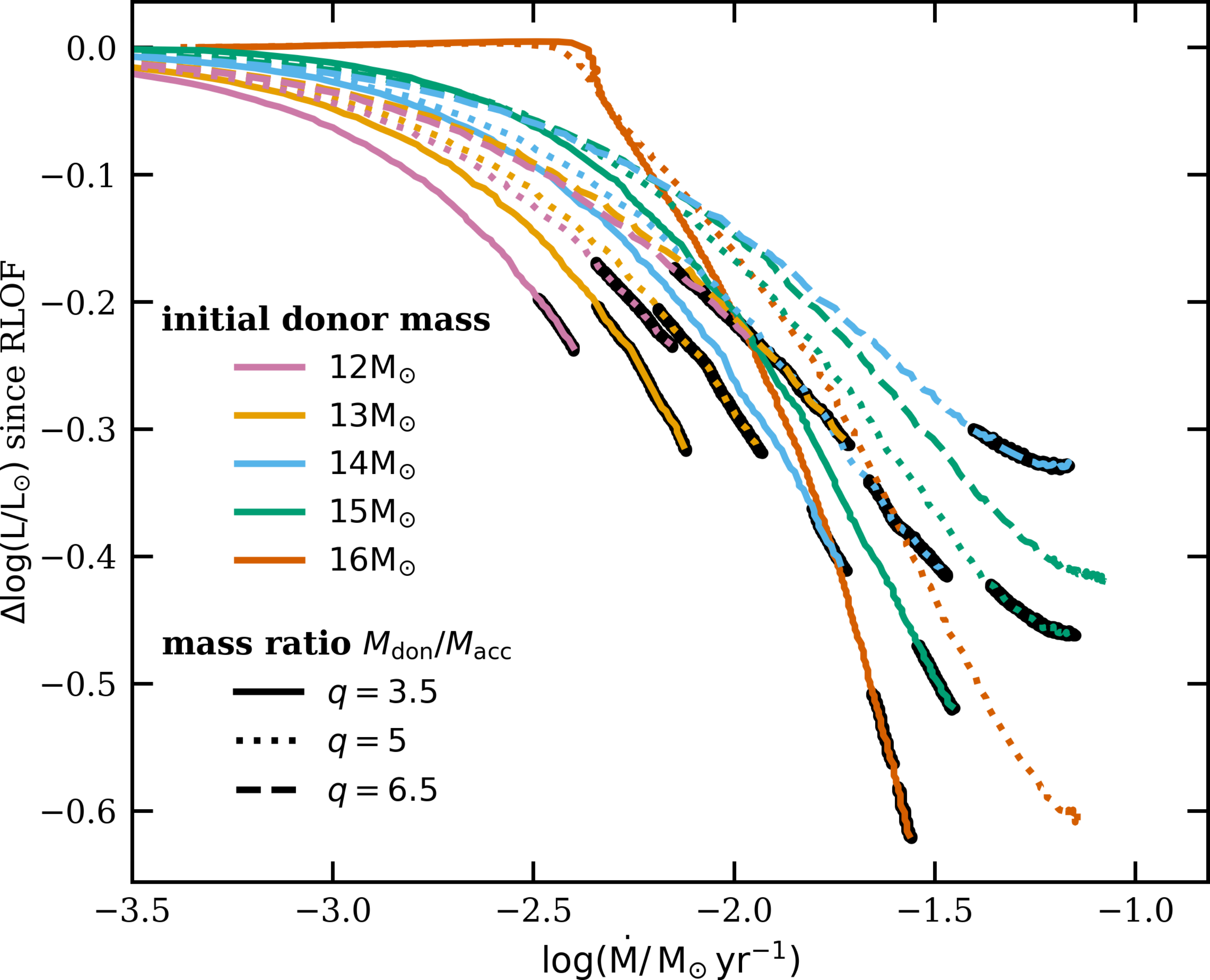} }
\caption{Relation between the current mass transfer rate in a binary and the decrease in the donor luminosity with respect to the moment of Roche-lobe overflow. The higher the rate with which the donor looses mass, the more energy is absorbed in an expanding envelope and, as a result, the lower is the surface luminosity. This general relation is degenerated with factors such as the mass ratio.}
\label{fig:deltaL_Mdot}
\end{figure}

Ideally, one would want to narrow down the range of possible initial donor masses, $M_{\rm don} = 12-16$\,\Msun, obtained from binary models. This, however, is not straightforward due to challenges in the modeling of interacting binaries with high mass-transfer rates ($\dot{M}$). As shown in the upper panel of Fig.~\ref{fig:Mdot_time}, donors of different initial masses are consistent with the locus of the \objname progenitor for different values of $\dot{M}$ (with ${\rm log}(\dot{M} / M_{\odot} \, {\rm yr}^{-1})$ ranging from $-2.4$ to $-1.2$). 
 This is because, in general, the higher the value of $\dot{M}$, the larger the decrease in the luminosity of the mass-losing donor (see Fig.~\ref{fig:deltaL_Mdot}). The details of this relation depend also on the current structure of the envelope and hence the amount of mass that has already been lost from it, which is sensitive to the assumed mass ratio. This explains the degeneracy between $M_{\rm don}$, $\dot{M}$, and $q$ among the binary models that are consistent with the progenitor, as shown in Fig.~\ref{fig:deltaL_Mdot}.
 
 It is difficult to say which value of $\dot{M}$ is consistent with the fact that only 10$-$13\,yr later the binary coalesced and produced a LRN event. The onset of a CE phase and a runaway dynamical inspiral is thought to be preceded by a stage of extensive loss of mass and angular momentum through L2 outflows, which decrease the separation to roughly the size of the donor's envelope \citep[e.g.,][]{Pejcha2016,PejchaMetzger2017ApJ}. This process can take a few hundred dynamical timescales or several years, during which $\sim$15\% of the companion mass needs to be ejected through L2 to provide sufficient orbital shrinkage \citep{MacLeod2017ApJ}.
 
 Thus, the progenitor system observed 10$-$13\,yr ago was most likely on the verge of experiencing significant L2 mass loss.  Such outflows could be triggered due to the donor star overflowing its Roche lobe up to the L2/L3 potential. The critical mass-transfer rate at which that happens is largely uncertain due to limitations of mass-transfer schemes in 1D stellar codes \citep{Pavlovskii2015}. Alternatively, significant L2 outflows may be triggered because of the companion star being driven out of thermal equilibrium and expanding significantly \citep[e.g.,][]{Benson1970,Neo1977}, 
 which could lead to it filling its Roche-lobe, a contact-binary stage \citep{Pols1994,Wellstein2001}, and potentially CE evolution \citep{deMink2007,Marchant2016}. Details of this process remain largely uncertain, 
partly because of the complicated gas dynamics in semi-detached binaries \citep{Lubow1975}, poorly constrained specific entropy of the accreted material \citep{Shu1981}, and unknown efficiency of accretion by stars rotating near their breakup limit \citep{Popham1991}. 
Because of these uncertainties, in the current study we are unable to exclude any of the binary model solutions based on the mass-transfer rate 10$-$13\,yr prior to the LRN event. We note that once significant L2 outflows commence, the mass-transfer rate is expected to quickly increase (see also Sect.~\ref{sec:sph}), changing the evolution of $\dot{M}$ in  Fig.~\ref{fig:Mdot_time} to a much steeper rise with time. \\

Alternatively, the mass transfer in the progenitor of \objname was still yet to commence 10$-$13\,yr before the LRN event. 
Such a scenario may be possible if the donor star is shrinking with respect to its Roche lobe on the adiabatic timescale.
For this to happen, the mass ratio between the binary components has to be large enough
(i.e., $q > q_{\rm crit;ad}$ where $q = M_{\rm don} / M_{\rm acc}$).
In the case of radiative-envelope supergiant donors such as the progenitor of \objname, the critical mass ratio $q_{\rm crit}$ derived
from adiabatic mass-loss sequences is $q_{\rm crit;ad} \approx $5$-$10 \citep{Ge2015} but increases to possibly even $q_{\rm crit;ad} > 20$ when
thermal relaxation of the outer envelope layers is taken into account \citep{Ge2020}.

The onset of the CE phase might also be very rapid when driven by Darwin (tidal) instability \citep{Darwin01011879}. 
The condition for tidal instability is based on the ratio between momentum of inertia of the orbit and of the donor star to be
$I_{\rm orb} / I_{\rm don} \lesssim 3$ \citep{Eggleton2001}. This condition can be rewritten into an expression for the critical 
separation below which the instability happens $a_{\rm crit} = R_{\rm don} \sqrt{3\eta_{\rm don}(1+q)}$ 
where the parameter $\eta_{\rm don} = I_{\rm don} M_{\rm don}^{-1} R_{\rm don}^{-2}$ is related to the internal 
structure of the donor \citep{MacLeod2017ApJ}. Using single MESA models (with same assumptions as in the binary MESA models described in Sect.~\ref{sec:binary_mesa_models}), we find that 10$-$12\,\Msun yellow supergiants 
consistent with the position of the \objname progenitor are characterized by $\eta_{\rm don} \approx $\,0.013$-$0.014. 
For such values of $\eta_{\rm don}$, the condition of the critical separation $a_{\rm crit}$ being equal to the RLOF separation 
requires extreme mass ratio values $q_{\rm crit;Darwin} \approx$\,45$-$50, which would correspond to companion stars of $\lesssim 0.2$\,\Msun. 

In summary, binary-evolution models suggest
a several-hundred year long phase of essentially stable mass transfer prior to the LRN event and yield a significantly higher progenitor mass (12$-$16\,\Msun) compared to single-star models ($\sim$11\,\Msun). Additional constraints are needed to further narrow down the progenitor mass range. Such constraints could come from clues on the stellar cluster environment of \objname (see Sect.~\ref{sec:disc_cluster}) or, in future work, from detailed modeling of the dust content of the progenitor system (see Sect.~\ref{sec:disc_dust_content}). Excitingly, any independent constrains on the progenitor mass or age, combined with binary-evolution models, would be a unique probe of the elusive conditions when rapid mass transfer leads to CE evolution, potentially helping to address some of the open questions outlined in this section.

An alternative scenario in which the progenitor of \objname was still in a detached state 10\,yr before the merger requires extremely steep mass ratios $\gtrsim$20. This seems unlikely, given that such systems are relatively rare \citep{Moe2017ApJS}, and this is also disfavored
based on the estimated amount of mass $M_{\rm ej}$ that was ejected from the system during the dynamical inspiral (see Sect.~\ref{sec:ce_prescription}).

\subsection{On the dust content in the progenitor system} \label{sec:disc_dust_content}

Independent constraints on the binary-evolution scenario may come from limits on the dust content of the progenitor system, obtained from the combination of archival \HST\/ photometry and \textit{SST\/} nondetections (Sect.~\ref{sec:progenitor_modelling}).
Based on a simple model of an optically thin and isothermal dust configuration, we find an upper limit on the mass of the dust at ${\rm log}(M_d / M_{\odot}) = -3.0$ if the dust temperature is $T_d = 250$\,K and much lower $M_d$ limits if $T_d$ is higher (e.g., ${\rm log}(M_d / M_{\odot}) = -6.0$ for $T_d = 1000$\,K, see Fig.~\ref{fig:dust_progenitor}). This amount of dust is at least a few times lower than the mass of metals $M_Z$ (i.e., future dust) that is transferred through L1 in our binary models for the progenitor, as shown in the lower panel in Fig.~\ref{fig:Mdot_time}, where $M_Z$ always reaches values $\gtrsim 2 \times 10^{-3}$\,\Msun.
It is expected that most of $M_Z$ will be ejected from the system and likely form dust \citep[for a discussion on accretion efficiency in case B mass transfer see][]{deMink2013,Vinciguerra2020}, seemingly in tension with the upper limits on the dust mass derived from the \textit{SST} data. Similarly, the mass-transfer rate through L1 is several orders of magnitude larger than the limit on optically thin dusty wind of $\lesssim 10^{-7}\,M_\odot\,\text{yr}^{-1}$ (Sect.~\ref{sec:progenitor_modelling}). 

There are two possible explanations why we did not detect dust in the progenitor, despite theoretical models suggesting copious mass loss. The first possible explanation is that the dust forms at much larger radii and lower temperatures, where it would radiate primarily in mid-IR and far-IR, where we do not have data. This could happen because of internal shocks in the outflow resetting grain growth by collisions or by UV radiation. Quantifying this possibility would require a much more sophisticated calculation beyond the scope of this work. The second possible explanation is that dusty outflow is radiatively inefficient. If we consider an equatorially concentrated outflow with velocity $v$, where radiation escapes perpendicular to the disk, the ratio of diffusion to expansion time is approximately
\begin{equation}
\tau \frac{v}{c} = \frac{\kappa X_\text{d} \dot{M}}{4\pi R_\text{form} c} \approx 0.6 \left(\frac{\kappa}{10^{3}\,\text{cm}^2\,\text{g}^{-1}} \right)\left(\frac{\dot{M}}{10^{-2}\,M_\odot\,\text{yr}^{-1}} \right)\left(\frac{R_\text{form}}{9\,\text{AU}}\right)^{-1},
\label{eq:dust_thick}
\end{equation}
where we assumed $X_\text{d} = 0.005$ and that the main contribution to emission comes from around the dust-formation radius, which is $R_\text{form} \approx 9$\,AU for the luminosity of \objname progenitor \citep{kochanek11}. For the theoretically expected $\dot{M}\approx 10^{-2}\,M_\odot\,\text{yr}^{-1}$, the grains grow to around $1\,\mu$m \citep{kochanek11}, which implies that the opacity of silicates at $5\,\mu$m is $\kappa \approx 10^{3}\,\text{cm}^2\,\text{g}^{-1}$. If $\tau v/c$ is around unity or higher, the outflow loses most of its internal energy to adiabatic expansion before it can be radiated. Given the scalings in Eq.~(\ref{eq:dust_thick}), the pre-merger binary mass loss in \objname could plausible be radiatively inefficient. This suggests that the theoretically predicted $\dot{M}$ could still be compatible with the \textit{SST\/} upper limits presented in Figure~\ref{fig:dust_progenitor}.

\subsection{Cluster environment of \objname{}}
\label{sec:disc_cluster}

\begin{figure*}[ht]
\hspace{-0.6cm}
\subfigure[]{\includegraphics[width=0.45\textwidth]{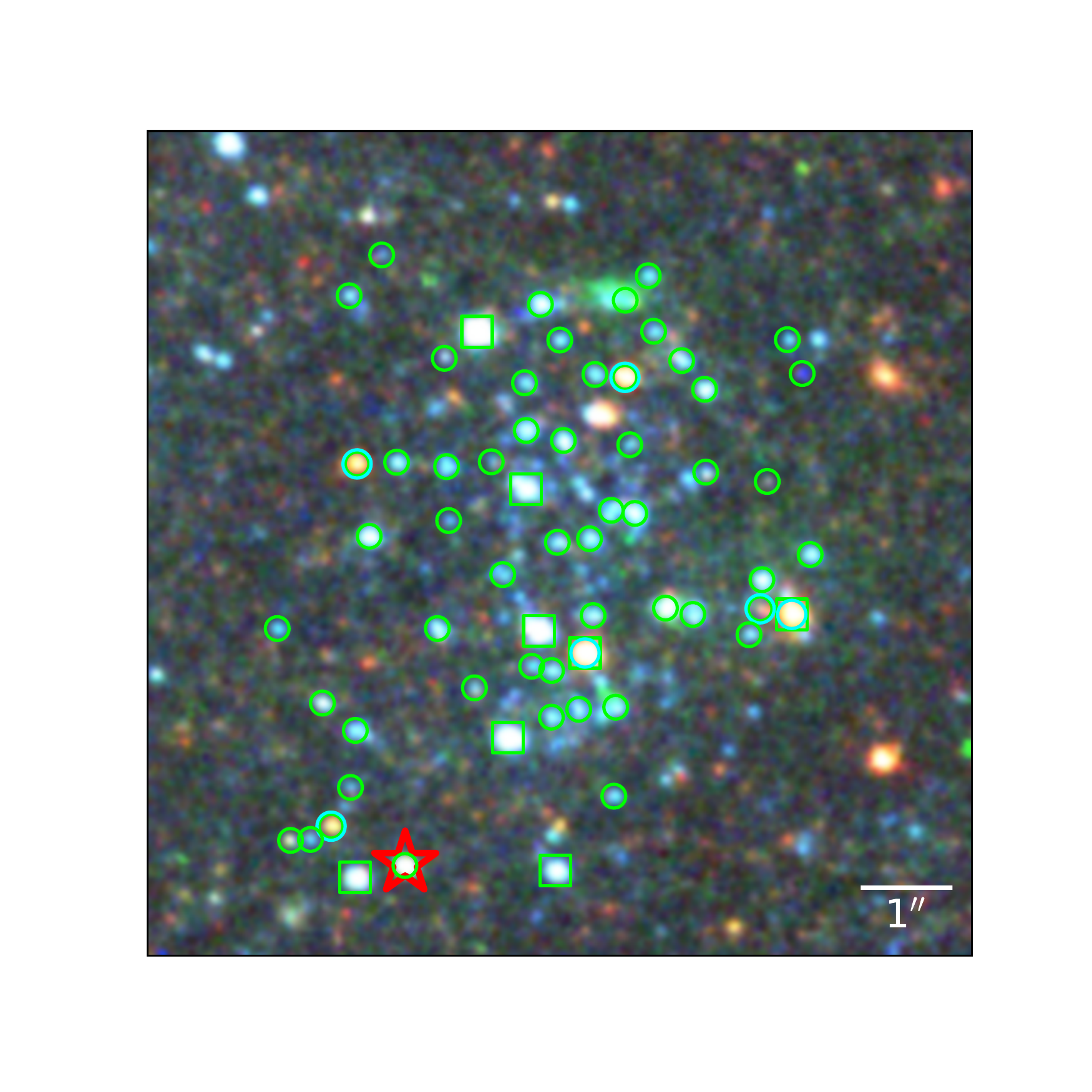}}
\hspace{-0.7cm}
\subfigure[]{\includegraphics[width=0.6\textwidth]{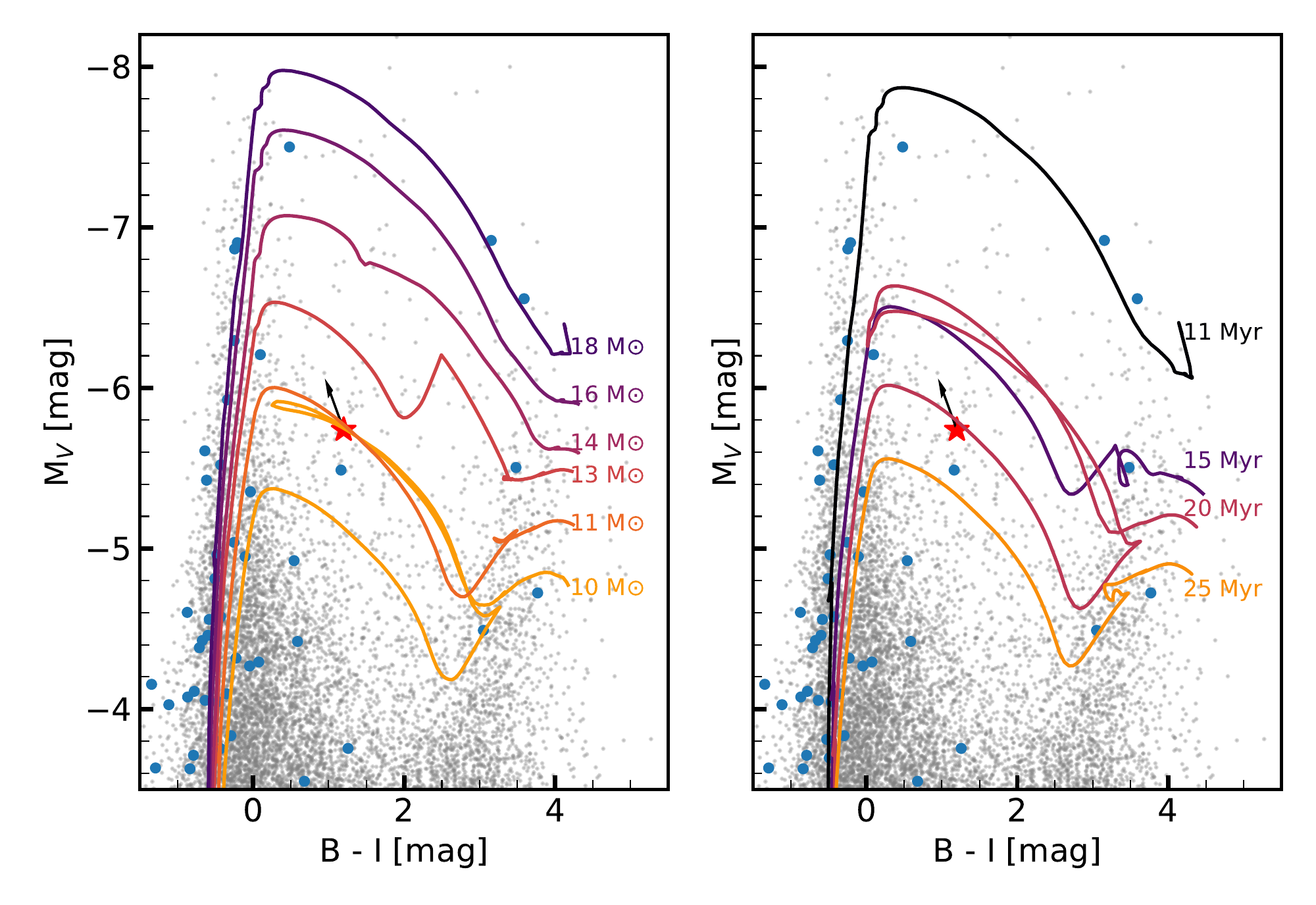}} 
\caption{Left: Cluster environment associated with the progenitor of \objname. Orientation north is up and east is left. The stars included in the analysis (cluster components) are marked with lime circles. The ones brighter than the progenitor in the HST $V$-band are marked with a square. The progenitor is marked with a red star. Middle: Colour-magnitude diagram in B$-$I (F435W $-$ F814W) bands for HST simulated magnitudes from MESA as compared to the progenitor magnitude. The initial mass for each model is labeled in the plot. Right: Same diagram, but showing the isochrones for different stellar ages. In both plots, the location of the \objname progenitor is marked with a red star. The black arrow indicates the location of the progenitor when corrected for A$_V=$0.2. The cluster components are shown in blue, and the rest of field stars in the same HST pointing are shown in faint gray.}
\label{fig:cluster_environment}
\end{figure*}

A closer analysis of the \objname progenitor site in NGC 45 suggests that the system was a likely member of a young stellar cluster, as shown in Fig.~\ref{fig:cluster_environment}. The observed group of stars has an angular radius of $\sim$2$''$, which at the distance of NGC 45 represents a radial extent of $\sim$64\,pc, consistent with an OB association. Assuming a similar formation time, the evolved primary progenitor of \objname would necessarily have been one of the most massive stars in the group, as other more massive components would have already ended their lives as core-collapse supernovae. However, the colour-magnitude diagram in Fig.~\ref{fig:cluster_environment} shows that potentially the cluster contains stars as massive as 16$-$18\,\Msun, with ages nearly half of the possible progenitors (as discussed in Sect.\ref{sec:disc_binary}, binary models predict ages between 12 and 18.5\,Myr for 16\,\Msun and 12\,\Msun donors respectively). Simultaneously, there are also RSGs of $\sim$10\,\Msun, which are $>7$\,Myr older than the progenitor.

This apparent discrepancy can easily be explained by the limited data available on the cluster. On one hand, the lack of radial velocities or proper motions limits our ability to distinguish between real cluster members and foreground or background contamination. On the other hand, unresolved binaries and unknown extinction around each star can also play a role in the location of each source in the diagram. Finally, the cluster may also have an intrinsic spread in age, and our initial assumption of co-evolution does not hold.

Despite the mentioned caveats, there seem to be two main populations in the cluster, with ages of $\sim$10 and 20\,Myrs. This is not unexpected, as blue stragglers from stellar mergers have been consistently observed in clusters, re-populating the main sequence above the main sequence turn-off point. In particular, in young ($\sim$5\,Myr) open clusters, their fraction near the turn-off may be as high as 30\% \citep{Schneider2015ApJ}. Observations of young clusters in the S~Doradus region in the LMC have also shown that the fraction for their analogous evolved counterparts, the red stragglers, can be as high as $\sim$55\% in a coeval cluster \citep{Britavskiy2019AA}.

Given that the progenitor is unlikely to be a merger product itself (unless it was originally a triple system), it likely belongs to the cluster's lower-mass population. Hence, we can use the observed RSGs to narrow down its mass to be in the 10$-$13\,\Msun regime, which is consistent with the 11$-$16\,\Msun mass range derived using both single and binary stellar-evolution models.

\section{Summary and conclusions}

In this work, we have presented the results of our photometric and spectroscopic follow-up campaign of the \objname LRN in the optical and IR wavelengths. We also modeled its progenitor system, which was observed by \HST\/ $\sim$10\,years before the outburst and, for the first time, determined the evolutionary stage of its progenitor system using binary stellar-evolution models.

\objname had a peak absolute magnitude of $M_r =-10.97\pm 0.11$\,mag, between those of V838\,Mon and M101\,OT2015-1, and comparable to AT\,2020hat. The duration of its $r$-band plateau of $41\pm5$\,days is in better agreement with lower-luminosity transients, such as M31-LRN2015 and AT\,2019zhd, both discovered in M31. 
Similarly to AT\,2020hat and other fainter LRNe, the early-time spectra of \objname were already marked by a cool continuum of $\sim$3000\,K and strong absorption lines of \io{Fe}{} and low-ionization elements. The \halpha lines also show the characteristic blue-shifted emission with an average FWHM of $\sim$500\,\kms and a narrow absorption component on top. Toward the end of the plateau, the appearance of strong \io{TiO}{} and \io{VO}{} molecular lines also matches previous LRNe observations. Our NIR spectra taken at 103\,days after discovery show an enlarged cool star, corresponding to a M8.5\,II type, analogous to AGB stars in the LMC. The progressive cooling and rapid creation of dust in the remnant is also observed in the \textit{NEOWISE} MIR data, which show that, although the object is not visible at NIR wavelengths, its emission in the MIR had increased 1.5\,years after the outburst.

Using MESA binary stellar-evolution models, we showed that the progenitor primary star is in the 12$-$16\,\Msun range, which is 9$-$45\% more massive than determined from single stellar-evolution models alone. We propose that the system was likely in a prolonged stage of semi-stable thermal-timescale mass transfer, with mass-transfer rates of $\dot{M} \sim 10^{-2} \, M_\odot \rm{yr}^{-1}$, allowing the primary to lose several $M_{\odot}$ before the dynamical onset of the CE. Surprisingly, this mass is not detected as an MIR dust excess in the \textit{SST} data of the system in quiescence. For optically thin warm (1500\,K $-$ 250\,K) dust, we place constraints of $10^{-8} - 10^{-3}$\,\Msun on the maximum dust mass present around the donor 10$-$13\,years before the outburst. This shows that the outflows from the primary were likely radiatively inefficient, or that the dust formation occurred at a larger radius, where temperatures are colder. We also suggest the onset of the dynamical instability was initiated by quick loss of angular momentum, caused by increasingly high L2/L3 mass loss, starting within the last few years before the outburst.

The analysis of the primary envelope's structure and the system's orbital energy support a partial ejection of the binary CE, likely within the range of 0.15$-$0.5\,\Msun. This mass is also in agreement with ejecta masses derived from modeling the energetics of the LRN outburst with both scaled supernova Type-II P (following \citet{Ivanova2013}) and shock-powered models \citep{MetzgerPejcha2017MNRAS}. Provided this mass is only a fraction of the total mass of the primary's envelope at the time of dynamical onset, we confirm that the LRN outburst \objname is related to a stellar-merger event.

Our results show that the combination of observations of progenitor systems and binary stellar-evolution models is a powerful tool to explore the conditions that may drive binary stars to unstable mass transfer and quick coalescence, which is critical to improve our understanding of the rapid mass-transfer evolution and the CE phase in binary systems. Future surveys, such as the Rubin Observatory Legacy Survey of Space and Time \citep[LSST;][]{LSST2009arXiv} are expected to discover $20-750$ LRNe per year \citep{Howitt2020MNRAS}. Although only the closest host galaxies are expected to have archival high resolution data ---critical to identify the progenitor systems in quiescence---, the depth of the survey will be ideal to exquisitely sample the precursor emission from the binary years before the LRN event, providing unique observational constraints on the intensive mass loss that drives the system to coalescence. Increasingly large samples of LRNe with detected progenitors and precursors will provide a rich opportunity to address the mass-transfer stability question from a statistical perspective, which will have an impact on binary population-synthesis models and hopefully improve our understanding on how binary stars evolve.

\section{Acknowledgements} \label{sec:acknowledgements}

\begin{small}
The authors would like to thank the members of the Stellar Evolution group at Radboud University, Tomasz {Kami{\'n}ski} and Morgan MacLeod for useful discussions; to S{\o}ren Larsen and Esteban Silva-Villa for sharing their photometric catalogs.
This work is part of the research programme VENI, with project number 016.192.277, which is (partly) financed by the Netherlands Organisation for Scientific Research (NWO).
This work was supported by the GROWTH project funded by the National Science Foundation (NSF) under grant AST-1545949. 
The research of OP has been supported by Horizon 2020 ERC Starting Grant `Cat-In-hAT' (grant agreement no. 803158).
RDG was supported, in part, by the United States Air Force.
J.K. acknowledges support from the Netherlands
Organisation for Scientific Research (NWO). 

Some of the data presented herein were obtained at the W. M. Keck Observatory, which is operated as a scientific partnership among the California Institute of Technology, the University of California, and the National Aeronautics and Space Administration (NASA); the observatory was made possible by the generous financial support of the W. M. Keck Foundation.
We acknowledge ESA Gaia, DPAC and the Photometric Science Alerts Team (http://gsaweb.ast.cam.ac.uk/alerts).
This work has made use of data from the Asteroid Terrestrial-impact Last Alert System (ATLAS) project. ATLAS is primarily funded to search for near earth asteroids through NASA grants NN12AR55G, 80NSSC18K0284, and 80NSSC18K1575; byproducts of the NEO search include images and catalogs from the survey area.  The ATLAS science products have been made possible through the contributions of the University of Hawaii Institute for Astronomy, the Queen's University Belfast, the Space Telescope Science Institute, and the South African Astronomical Observatory.
This publication makes use of data products from the Two Micron All Sky Survey, which is a joint project of the University of Massachusetts and the Infrared Processing and Analysis Center/California Institute of Technology, funded by NASA and the NSF. 
This publication makes use of data products from the Wide-field Infrared Survey Explorer, which is a joint project of the University of California, Los Angeles, and the Jet Propulsion Laboratory/California Institute of Technology, funded by NASA.
This work makes use of observations from the Las Cumbres Observatory global telescope network.
Based on observations made with the NASA/ESA Hubble Space Telescope, and obtained from the Hubble Legacy Archive, which is a collaboration between the Space Telescope Science Institute (STScI/NASA), the Space Telescope European Coordinating Facility (ST-ECF/ESAC/ESA) and the Canadian Astronomy Data Centre (CADC/NRC/CSA).

\end{small}

\bibliographystyle{aa}
\bibliography{main}

\end{document}